%% file: main.tex
\documentclass[sigconf,balance=false]{acmart}

\usepackage{popets}
\setcopyright{popets}
\copyrightyear{YYYY}
\acmYear{YYYY}
\acmVolume{YYYY}
\acmNumber{X}
\acmDOI{XXXXXXX.XXXXXXX}
\acmISBN{}
\acmConference{Proceedings on Privacy Enhancing Technologies}
\usepackage{tikz}
\usepackage{amsmath}
\usepackage{filecontents}
\usepackage{blindtext}
\usepackage{subcaption}
\usepackage{caption}
\usepackage{courier}
\usepackage{newfloat}
\usepackage{comment}
\usepackage{xspace}
\usepackage{foreign}
\usepackage{booktabs}
\usepackage{balance}
\usepackage{natbib}
\usepackage{graphicx}
\usepackage{algorithm}
\usepackage{algorithmic}
\usepackage{listings}
\usepackage{color}
\usepackage{amsmath}
\usepackage{amstext}
\usepackage{multirow}
\usepackage{graphicx}
\usepackage{lipsum}
\usepackage{pifont}
\usepackage{todonotes}
\usepackage{enumitem}
\usepackage{tcolorbox}
\usepackage{seqsplit}

\def\eg{\emph{e.g.}\xspace}
\def\ie{\emph{i.e.}\xspace}
\def\etc{\emph{etc.}\xspace}

\newcommand{\one}{({\em i}\/)\xspace}
\newcommand{\two}{({\em ii}\/)\xspace}
\newcommand{\three}{({\em iii}\/)\xspace}
\newcommand{\four}{({\em iv}\/)\xspace}

\newcommand{\pb}[1]{\vspace{0.75ex}\noindent{\bf \em #1}\hspace*{.3em}}

\newcommand{\takehomebox}[1]{%
  \begin{tcolorbox}[colback=white, colframe=black!75!black, rounded corners=north, boxrule=0.5pt, boxsep=0.1mm, width=\columnwidth]
    #1
  \end{tcolorbox}
}

\begin{document}

\input{includes/abbreviations}
\input{includes/data}

\title{Measuring the Accuracy and Effectiveness of PII Removal Services}

\author{Jiahui He}
\affiliation{%
  \institution{The Hong Kong University of Science and Technology (Guangzhou)}
  \city{}
  \state{}
  \country{}}
\email{jhe976@connect.hkust-gz.edu.cn}

\author{Peter Snyder}
\affiliation{%
  \institution{Brave Software Inc}
  \city{}
  \state{}
  \country{}}
\email{pes@brave.com}

\author{Hamed Haddadi}
\affiliation{%
  \institution{Imperial College London \& Brave Software Inc}
  \city{}
  \state{}
  \country{}}
\email{h.haddadi@imperial.ac.uk}

\author{Fabián E. Bustamante}
\affiliation{%
  \institution{Northwestern University}
  \city{}
  \state{}
  \country{}}
\email{fabianb@northwestern.edu}

\author{Gareth Tyson}
\affiliation{%
  \institution{The Hong Kong University of Science and Technology (Guangzhou)}
  \city{}
  \state{}
  \country{}}
\email{gtyson@ust.hk}

\renewcommand{\shortauthors}{He et al.}

\begin{abstract}
\input{Section/0.Abstract}
\end{abstract}

\keywords{data brokers, personal information, user privacy, user study}

\maketitle

\input{Section/1.Introduction}
\input{Section/2.Motivation_Background}
\input{Section/3.Data_Collection}

\input{Section/4.Services_Characterization}
\input{Section/6.Crowdsourcing}
\input{Section/7.Discussion}
\input{Section/8.Related_Work}
\input{Section/9.Conclusion}

\begin{acks}
This work was supported in part by the Guangzhou Science and Technology Bureau (2024A03J0684), Guangdong provincial project 2023QN10X048, the Guangzhou Municipal Key Laboratory on Future Networked Systems (2024A03J0623), the Guangdong Provincial Key Lab of Integrated Communication, Sensing and Computation for Ubiquitous Internet of Things (No.2023B1212010007), the Guangzhou Municipal Science and Technology Project (2023A03J0011), Guangdong provincial project (2023ZT10X009).
Haddadi wishes to acknowledge funding from UKRI OpenPlus Fellowship EP/W005271/1.
\end{acks}

\balance
\bibliographystyle{plain}
\bibliography{references}

\appendix

\section{APPENDIX}

\subsection{PII Removal Services Google Trends}
\label{appendix:a3}

\begin{figure}[h]
     \centering
     \includegraphics[width=.9\linewidth]{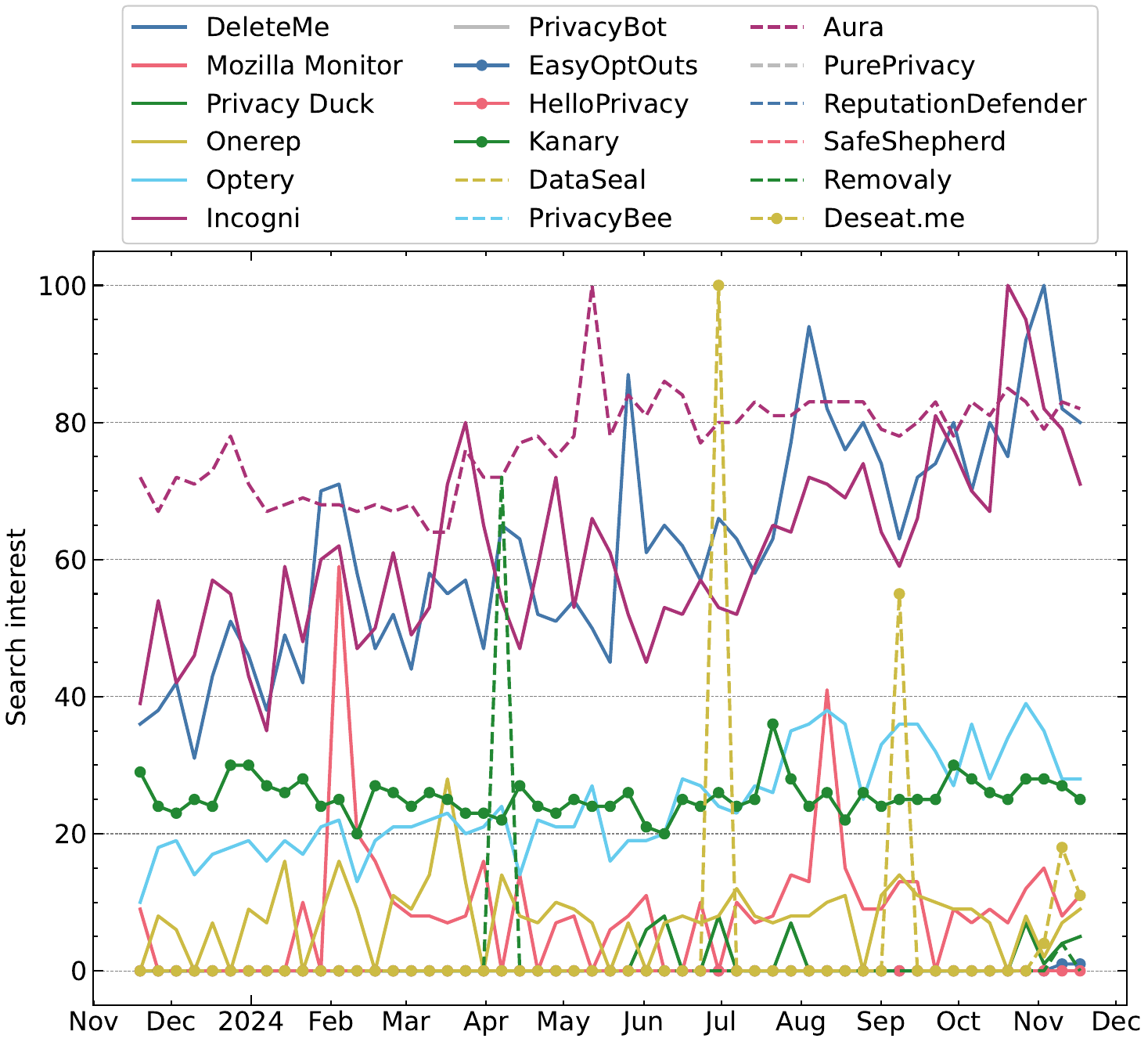}
     \caption{Weekly average of Google search interest results for 18 PII removal services' name from Nov 2023 to Dec 2024.}
     \label{fig:google_trend_split}
\end{figure}

Figure \ref{fig:google_trend_split} shows the Google search interest for the 18 PII removal service names separately.

\subsection{eTLD+1 Group}
\label{appendix:a4}

\begin{table*}[]
\resizebox{2\columnwidth}{!}{%
\begin{tabular}{ll}
\hline
\textbf{Domain}  & \textbf{Group}                                              \\ \hline
33across.com            & 33across.com, udp.33across.com                              \\
ancestry.com            & ancestry.com, search.ancestry.com                           \\
atdata.com              & atdata.com, instantdata.atdata.com                          \\
bigdbm.com              & bigdbm.com, optout.bigdbm.com                               \\
careerbuilder.com & careerbuilder.com, hiring.careerbuilder.com, screen.careerbuilder.com                                                        \\
cataloxy.us             & ct-state.cataloxy.us, ma-framingham.cataloxy.us             \\
classmates.com          & classmates.com, help.classmates.com                         \\
clearbit.com            & clearbit.com, preferences.clearbit.com                      \\
cmac.ws                 & cosmetics-stores.cmac.ws, nurses-and-midwives.cmac.ws       \\
criminalregistry.org    & criminalregistry.org, jeremy-koski.criminalregistry.org     \\
cybo.com                & cybo.com, halaman-kuning.cybo.com, yellowpages-hi.cybo.com  \\
dandb.com               & a.assets.dandb.com, dandb.com                               \\
dataxltd.com            & consumers.dataxltd.com, dataxltd.com                        \\
dateas.com              & dateas.com, m.dateas.com                                    \\
deepsync.com            & deepsync.com, privacy.deepsync.com                          \\
deluxe.com              & deluxe.com, fi.deluxe.com                                   \\
epsilon.com             & epsilon.com, legal.epsilon.com, us.epsilon.com              \\
equifax.com             & equifax.com, myprivacy.equifax.com, totalverify.equifax.com \\
fetcher.ai              & app.fetcher.ai, fetcher.ai                                  \\
findlaw.com             & caselaw.findlaw.com, lawyers.findlaw.com                    \\
getemail.io             & b2b.getemail.io, getemail.io                                \\
google.com              & docs.google.com, google.com, groups.google.com              \\
healthprovidersdata.com & healthprovidersdata.com, webmail.healthprovidersdata.com    \\
infofree.com            & infofree.com, profile.infofree.com                          \\
information.com         & information.com, searchportal.information.com               \\
infotracer.com          & infotracer.com, members.infotracer.com                      \\
jellyfish.com           & info.jellyfish.com, jellyfish.com                           \\
knowwho.com             & knowwho.com, kw1.knowwho.com                                \\
lead411.com             & app.lead411.com, lead411.com                                \\
lexisnexis.com    & consumer.risk.lexisnexis.com, lexisnexis.com, optout.lexisnexis.com, risk.lexisnexis.com                                     \\
michigancorporates.com  & en.michigancorporates.com, michigancorporates.com           \\
minervadata.xyz  & minervadata.xyz, realtors.minervadata.xyz           \\
monitorbase.com         & monitorbase.com, www.monitorbase.com                        \\
moodys.com              & ma.moodys.com, moodys.com                                   \\
moodysanalytics.com     & cre.moodysanalytics.com, pulse.moodysanalytics.com          \\
onetrust.com      & privacyportal-cdn.onetrust.com, privacyportal-eu-cdn.onetrust.com, privacyportal-eu.onetrust.com, privacyportal.onetrust.com \\
onlinesearches.com      & onlinesearches.com, publicrecords.onlinesearches.com        \\
oracle.com              & datacloudoptout.oracle.com, oracle.com                      \\
propublica.org          & projects.propublica.org, propublica.org                     \\
public-record.com       & consumer.public-record.com, mtis-consumer.public-record.com \\
searchsystems.net       & publicrecords.searchsystems.net, searchsystems.net          \\
selfie.systems          & 18ip.selfie.systems, selfie.systems                         \\
spglobal.com            & more.spglobal.com, spglobal.com                             \\
staterecords.org  & districtofcolumbia.staterecords.org, members.staterecords.org, oklahoma.staterecords.org, staterecords.org                   \\
targetsmart.com         & privacy.targetsmart.com, targetsmart.com                    \\
telephonedirectories.us & premium.telephonedirectories.us, telephonedirectories.us    \\
theknot.com             & registry.theknot.com, theknot.com                           \\
thomsonreuters.com      & legal.thomsonreuters.com, thomsonreuters.com                \\
uscourts.gov            & ohnb.uscourts.gov, tneb.uscourts.gov                        \\
verisk.com              & marketing.verisk.com, verisk.com                            \\
whitepages.com          & premium.whitepages.com, whitepages.com                      \\
yahoo.com               & local.yahoo.com, search.yahoo.com                           \\
yellowpages.com         & people.yellowpages.com, yellowpages.com                     \\
yp.ca                   & corporate.yp.ca, yp.ca                                      \\
zenprospect.com         & blog.zenprospect.com, zenprospect.com                       \\ \hline
\end{tabular}%
}
\caption{54 groups of data broker domains with the same eTLD+1.}
\label{table:public_suffix}
\end{table*}

Table \ref{table:public_suffix} shows 55 groups of data brokers with the same eTLD+1.

\subsection{Browser Plugin Implementation}
\label{appendix:a5}

To verify user subscriptions and the data being sent, we develop a browser plugin using JavaScript. 
Our backend (which receives data) runs on Google Cloud platform, which has two CPU cores and 1 GB of memory. This virtual machine has an external IP address, allowing participants to send data to that IP through the plugin.

After participants subscribe to the PII removal service, they must use the plugin to check whether their subscription is valid. 
This allows us to confirm that the participants have subscribed correctly. The participants must open the PII removal service, upon which the plugin automatically parses the page's HTML and sends this data (in JSON format) to the backend for validation. This data \emph{only} includes the participant's current subscription type and does not contain any personal information.

After 30 days of subscription, participants send us data regarding the removal progress through the browser plugin. When a participant opens the removal service's progress page (which includes all retrieved records and their removal statuses), the plugin automatically extracts the HTML from the current page. It then filters out the necessary data and sends this information (in JSON format) to our backend. Again, it is important to highlight that this data does not include any personal information about the participants. Instead, it only includes the removal status for a particular type of PII, and not the PII value itself (\eg "Age" rather than "35").
The extension is available on the Chrome Web Store,\footnote{\url{https://chromewebstore.google.com/detail/services-progress-result/odefebdiianlgejbdbfhpbkpmodmhpaj?hl=en-US&utm_source=ext_sidebar}} and the source code is available on GitHub.\footnote{\url{https://github.com/HHHeJiahui/Data-Broker-Extractor}}

\begin{figure*}
     \centering
     \includegraphics[width=0.6\linewidth]{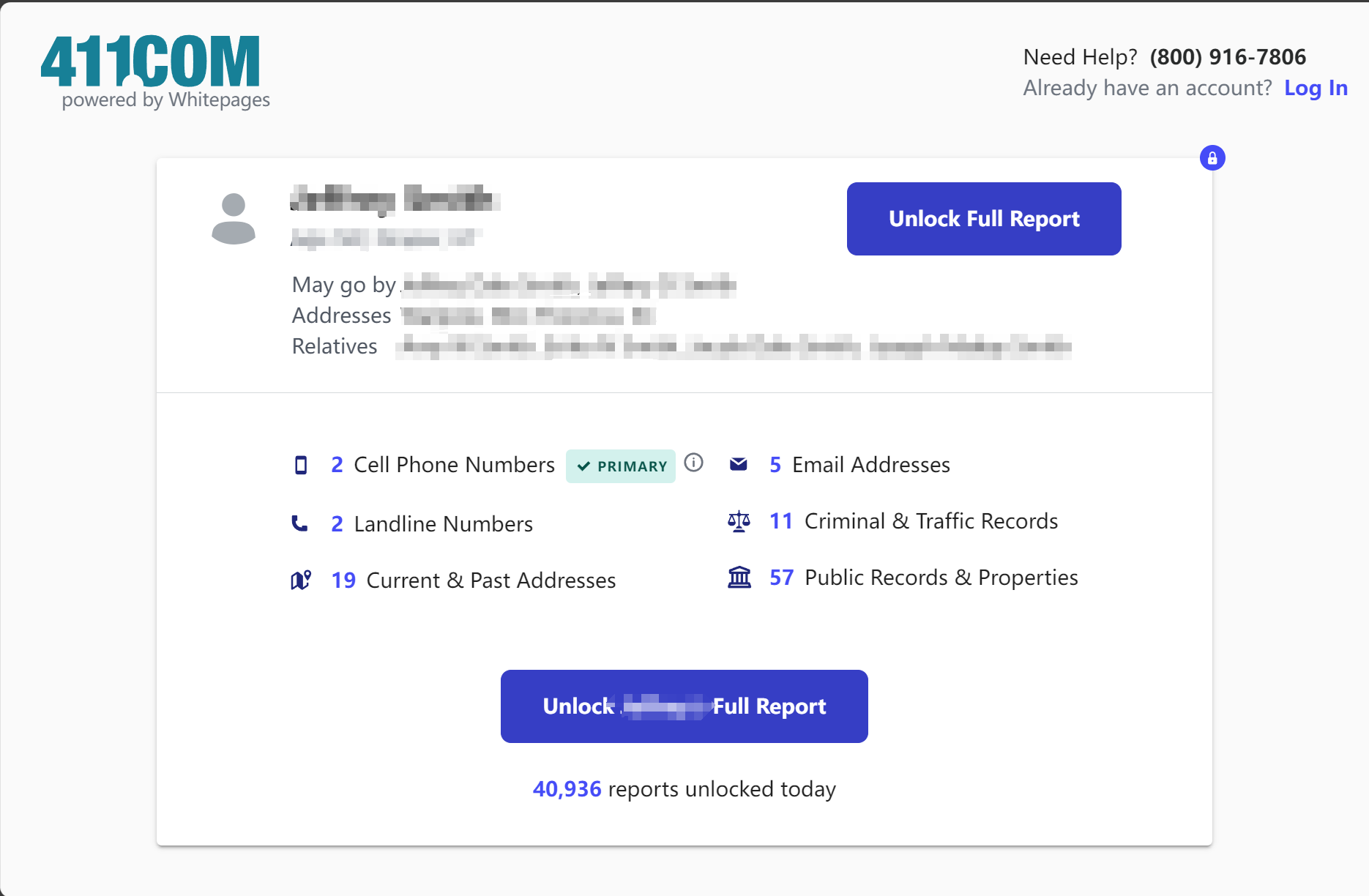}
     \caption{Screenshot of records returned by searching on the 411.com website (the private information has been mosaicked).}
     \label{fig:data_broker_search_result}
\end{figure*}

\subsection{Consent Form}
\label{appendix:a6}
\pb{Research Title}: A Study to evaluate the effectiveness of services that help individuals remove personal information from data broker databases.

\pb{Procedures}: If you agree to participate, you will be asked to subscribe to the specified data removal service and send us service removal progress data after one month via a browser extension.

\pb{Data Collection}: We commit not to collect any private data about you. We only collect removal progress data of services, such as when your personal information was removed from which data broker. The data collected will be used to assess the effectiveness of removals for each service.

\pb{Confidentiality}: Although the data we collect does not contain any personal, private data, all data will still be stored securely and only accessible to the research team. 

\pb{Voluntary Participation}: Participation in this study is entirely voluntary. You may choose not to participate or withdraw at any time, but please note that early withdrawal may result in the loss of the \$40 Amazon shopping card bonuses. If you need to quit the experiment, please send an email to hejiahui14756@gmail.com.

\pb{IRB Approval}: This study has been reviewed and approved by the Institutional Review Board (IRB) of The Hong Kong University of Science and Technology (GZ). The IRB has determined that this study meets the ethical standards for research involving human subjects (HSP-2024-0023).

\pb{Contact Information}: If you have any questions or concerns about this study, please contact hejiahui14756@gmail.com

\pb{Consent}: By completing the google consent form, you acknowledge that you have read and understand the information above, and you agree to participate in this study.

\subsection{VirusTotal Category Verification Result}
\label{appendix:a7}

\begin{figure*}
     \centering
     \includegraphics[width=.6\linewidth]{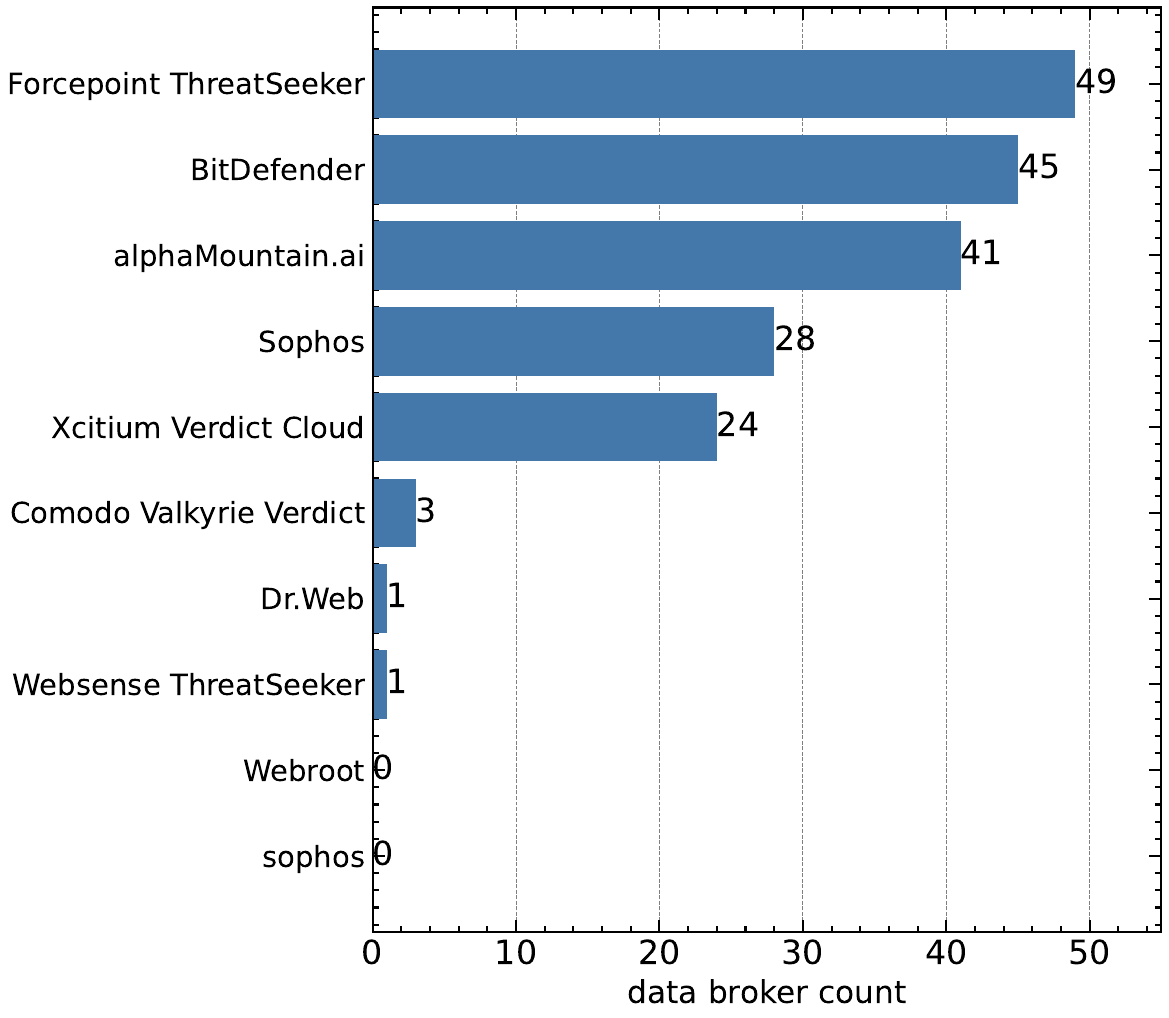}
     \caption{Categorization accuracy of domain categorization engines for 50 random data brokers. }
     \label{fig:different_category}
\end{figure*}

Figure \ref{fig:different_category} shows the categorization accuracy of several domain categorization engine for 50 random data broker domains. These engines were taken from VirusTotal. Accuracy is calculated by manually verifying the correctness. \emph{Forcepoint ThreatSeeker} has the highest number of valid classifications.

\end{document}

%% file: includes/abbreviations.tex
\newcommand{\DB}[1][]{data broker#1}

% Adding this bc everyone has a different opinion on "Web" vs "web"
\newcommand{\Web}{Web}

% Adding this bc of similar debates of "i.e.," vs "ie," vs "i.e." etc
\newcommand{\IE}{i.e.,}
\newcommand{\EG}{e.g.,}

%% file: includes/data.tex
\newcommand{\NumServicesBroad}{10}
\newcommand{\NumServicesBroadWord}{ten}
\newcommand{\NumServicesDetailed}{4}
\newcommand{\NumServicesDetailedWord}{four}
\newcommand{\NumDataBrokers}{2,024}

\newcommand{\NumParticipants}{71}
\newcommand{\NumParticipantsWord}{seventy-one}

% What percent of records were able to be deleted
\newcommand{\ServicesDeletionRate}{48.2\%}

% What pct of the records the services found were actually PII about the user
\newcommand{\ServicesAccuracyRate}{41.1\%}

%% file: Section/0.Abstract.tex
This paper presents the first large-scale empirical study of commercial personally identifiable information (PII) removal systems --- commercial services that claim to improve privacy by automating the removal of PII from data broker's databases. Popular examples of such services include \texttt{DeleteMe}, \texttt{Mozilla Monitor}, \texttt{Incogni}, among many others. The claims these services make may be very appealing to privacy-conscious Web users, but how effective these services actually are at improving privacy has not been investigated.
This work aims to improve our understanding of commercial PII removal services in multiple ways. First, we conduct a user study where participants purchase subscriptions from four popular PII removal services, and report \one~what PII the service find, \two~from which data brokers, \three~whether the service is able to have the information removed, and \four~whether the identified information actually is PII describing the participant. And second, by comparing the claims and promises the services makes (\eg which and how many data brokers each service claims to cover). We find that these services have significant accuracy and coverage issues that limit the usefulness of these services as a privacy-enhancing technology. For example, we find that the measured services are unable to remove the majority of the identified PII records from data broker's (48.2\% of the successfully removed found records) and that most records identified by these services are \emph{not} PII about the user (study participants found that only 41.1\% of records identified by these services were PII about themselves).

%% file: Section/1.Introduction.tex
%-------------------------------------------------------------------------------
\section{Introduction}
%-------------------------------------------------------------------------------
Personal Identifiable Information (PII) has become a core of the global information economy.
PII refers to any data that can be used to identify a specific individual, including information such as names, addresses, emails, phone numbers, and even biometric data like fingerprints \cite{rozenberg2012challenges}. 
With the continuous growth in the value of PII, the risk and frequency of data exposure has also increased \cite{fleury2022data, ayyagari2012exploratory, hall2018data}.
Such incidents not only violate personal privacy, but may also lead to identity theft, financial fraud, and long-term damage to the reputation of individuals and organizations.

In this digital ecosystem, \emph{data brokers} play an important and controversial role by treating PII as a commodity. They collect and trade vast amounts of PII --— often without individuals’ knowledge or consent —-- from sources such as public records, online activities, social media, and retail transactions \cite{crain2018limits, kuempel2016invisible, McAfee}. 
The data broker market, projected to reach \$382.16 billion by 2030 \cite{MMR}, fuels concerns about consent, ownership, and misuse of PII. For instance, the 2017 Equifax breach compromised the PII of 147 million people, highlighting the systemic risks posed by centralized repositories of sensitive data \cite{Equifax}. These risks underscore the urgent need for mechanisms to mitigate the unchecked proliferation of PII.

In response to the risks posed by the poorly regulated collection of personal PII by data brokers, various laws have been issued. F
GDPR, implemented in 2018, established strong data protection standards for EU citizens, requiring organizations to be transparent about data collection and usage \cite{li2019impact, albrecht2016gdpr, voigt2017eu, zaeem2020effect}. Similarly, the CCPA, implemented in 2020, grants California residents the right to know about their personal data, its collection purpose, and sharing practices, while also allowing them to opt-out of data sales \cite{bonta2022california, de2018guide, barrett2019eu}.
However, enforcement gaps and the opaque nature of data brokerage persist, leaving individuals with limited practical control over their PII \cite{kuempel2016invisible, dipersio2024selling}.

In order for individuals to better protect personal privacy, a new industry has emerged: \emph{PII removal services}.
These services act on behalf of individuals (typically for a small fee). They actively seek to remove users' PII from data brokers and other online platforms to reduce the risk of misuse of PII and data breaches. 
The process typically involves identifying which data brokers hold an individual's data, submitting formal removal requests, and monitoring for compliance. PII removal services often employ a combination of automated tools and manual processes to efficiently manage data removal requests.
By intervening in the data broker ecosystem, the PII removal service arguably strengthens individual's control over their own data and also contributes positively to the promotion of privacy protection.

Despite the growing role these PII removal services play in the privacy ecosystem, they remain critically understudied. To bridge this gap, we conduct the first large-scale empirical study on PII removal services. Based on an initial survey of 10 major PII removal services and the 2,024 data brokers they cover, we recruit 71 participants to use 4 different services. Through this, we study what PII the services discover on each data broker, alongside whether the service is able to (correctly) remove the PII. We explore the following research questions:

\begin{itemize}[leftmargin=*]
    \item \textbf{RQ1}: What are the \emph{characteristics} of PII removal services in terms of the data brokers they cover and the information they require to pursue data removals?
    
    \item \textbf{RQ2}: How effectively do these PII removal services \emph{discover} user records on data brokers, in terms of the number and accuracy of records retrieved?
    
    \item \textbf{RQ3}: What is the efficacy of the PII removal service in \emph{removing} the discovered records? Specifically, what is the percentage of records successfully removed from the data brokers, and what is the time required to achieve this?
\end{itemize}

Through studying these RQs, our findings include:

\begin{enumerate}[leftmargin=*]
    \item The data broker coverage of the PII removal services varies widely, with a low overlap between them (average Jaccard similarity of 0.21). There are only 10 data brokers common to all services, indicating they may target different industries or user groups. We find that 71.7\% of data brokers covered are not government-registered, showing a lack of regulation. ((\S\ref{subsec:service_characterization1})
    
    \item The removal services vary in terms of their requested user PII (required to facilitate the data removals). \textbf{Full Name}, \textbf{Email}, and \textbf{City Address} are mandatory across all of them. However, services with larger data broker coverage tend to request more PII from users during the subscription stage. (\S\ref{subsec:service_characterization2})

    \item Data brokers operate across various industries, predominantly in business (19.0\%) and information technology (16.0\%). However, aside from those in the economy and finance industry, most data brokers are not registered with the required government authorities, highlighting insufficient regulatory practices. (\S\ref{subsec:data_broker_characterization3})

    \item Removal services that boast of covering a larger number of brokers, do not necessarily have high success rates. Notably, \texttt{Kanary} has the worst performance, despite having the widest data broker coverage. In fact, it identifies the lowest number (average of 14.6) of records for the users from its broker coverage. This indicates that larger data broker coverage does not necessarily mean that PII can be successfully removed from all listed data brokers. (\S\ref{subsec:crowdsourcing_experiment1})

    \item The accuracy of the PII removal service in retrieving records is also low, with an average of 41.1\% of records being correctly linked to the participant (\ie where the removed record contains valid information about the participant). This means services are potentially removing a large number of records that do not belong to the subscribing user, but rather other people with similar PII (\eg same name, same date of birth \etc). (\S\ref{subsec:crowdsourcing_experiment2})

    \item We find that the PII removal services are only successful in removing an average of 48.2\% of the identified records per user. Even in cases where a service is successful in removing one user's record from a data broker, it may be unsuccessful in removing another user's record from the same broker. (\S\ref{subsec:crowdsourcing_experiment3})

\end{enumerate}

%% file: Section/2.Motivation_Background.tex
%-------------------------------------------------------------------------------
\section{Background}
\label{sec:motivation_background}
%-------------------------------------------------------------------------------

\pb{Data Brokers.}
Data brokers (aka information brokers or data vendors) operate in a complex ecosystem where PII is a valuable commodity. Data brokers collect, analyze, aggregate, and package up PII from various sources, including web crawls, public records, online activity, self-reported information and other data brokers \cite{crain2018limits, McAfee}. 
These packaged datasets are then re-sold to third parties, such as marketers, advertisers, financial institutions, and sometimes even government agencies, often for substantial profit. The scope of PII collected by data brokers includes, but is not limited to, name, email, physical address, date of birth, phone number, relatives, employment, health issues, social network connections.
For example, \url{411.com} (a well-known people search site) allows users to search for individuals by name and location, returning results that may include age, phone number, address, and email. For context, Figure \ref{fig:data_broker_search_result} in the Appendix shows a screenshot of the search results on \url{411.com}. 
However, for most brokers, viewing detailed information typically requires a subscription --- this makes it difficult for users to understand what data is stored about themselves. The data broker industry is lucrative and largely unregulated, raising concerns about privacy, consent, and the ethical implications of extensive data collection and dissemination.

\pb{Opt-Out Process.}
Given the potential for misuse of PII, the concept of ``opt-out'' has emerged as a means for individuals to control their PII. Opting-out involves requesting that an organization (\eg a data broker) refrains from collecting, selling, or sharing an individual's PII. However, this process can be complex and lacks standardization across the industry.
Some PII removal services publish opt-out guides for certain data brokers on their official websites, \eg DeleteMe,\footnote{\url{https://joindeleteme.com/blog/opt-out-guides/}} Optery\footnote{\url{https://www.optery.com/opt-out-guides/}} and Incogni.\footnote{\url{https://blog.incogni.com/opt-out-guides/}}
Typically, the opt-out process requires individuals to locate their personal information page on the data broker's site and submit an opt-out request through a series of forms or by sending emails, which are often not visibly displayed. Compounding this difficulty, the specific requirements and procedures vary dramatically across different data brokers, resulting in a fragmented system plagued by inconsistent standards and excessive complexity. Thus, it is time-consuming and complex for individuals to opt-out from multiple data brokers.

\pb{PII Removal Services.}
To address the challenge that opt-out is overly complex, PII removal services provide a simple portal designed to simplify the process. 
The PII removal service maintains its own data broker coverage list, indicating which brokers it can help users remove their PII from. To begin the process, users must submit their PII to the removal service (see \S\ref{subsec:service_characterization2}). The service then uses this information to search for the user’s records across its covered data brokers. Once identified, it automatically submits opt-out requests to those brokers on the user’s behalf, thereby reducing complexity.
However, the effectiveness of these opt-out procedures varies: some brokers may delete PII promptly, while others may even ignore such requests Currently, there is a lack of clear evidence evaluating the efficacy of these removal services and the behaviors of the data brokers they contact.

%% file: Section/3.Data_Collection.tex
%-------------------------------------------------------------------------------
\section{Data Collection \& Methodology}
\label{sec:data_collection}
%-------------------------------------------------------------------------------

\subsection{PII Removal Services}
\label{subsec:data_collection2}

\pb{Discovery of Removal Services.}
There are a wide variety of PII removal services available in-the-wild with different data broker coverage. To identify the most important/popular services, we start with Google Trends \cite{piasecki2018google}, searching for relevant topics about ``Data Broker" over the past year.
This gives us some well known services to use as \emph{seed} terms, which are: \texttt{Incogni} \cite{Incogni}, \texttt{Aura} \cite{Aura} and \texttt{DeleteMe} \cite{DeleteMe}.
We then explore more services by searching ``\emph{[seed]} similar website'', ``\emph{[seed]} alternatives'', ``\emph{[seed]} competitors'' on Google, and searching the same keywords on famous rating \& forum websites: Reddit\footnote{\url{https://www.reddit.com/}} and Quora.\footnote{\url{https://www.quora.com/}}

\begin{figure}
     \centering
     \includegraphics[width=0.8\linewidth]{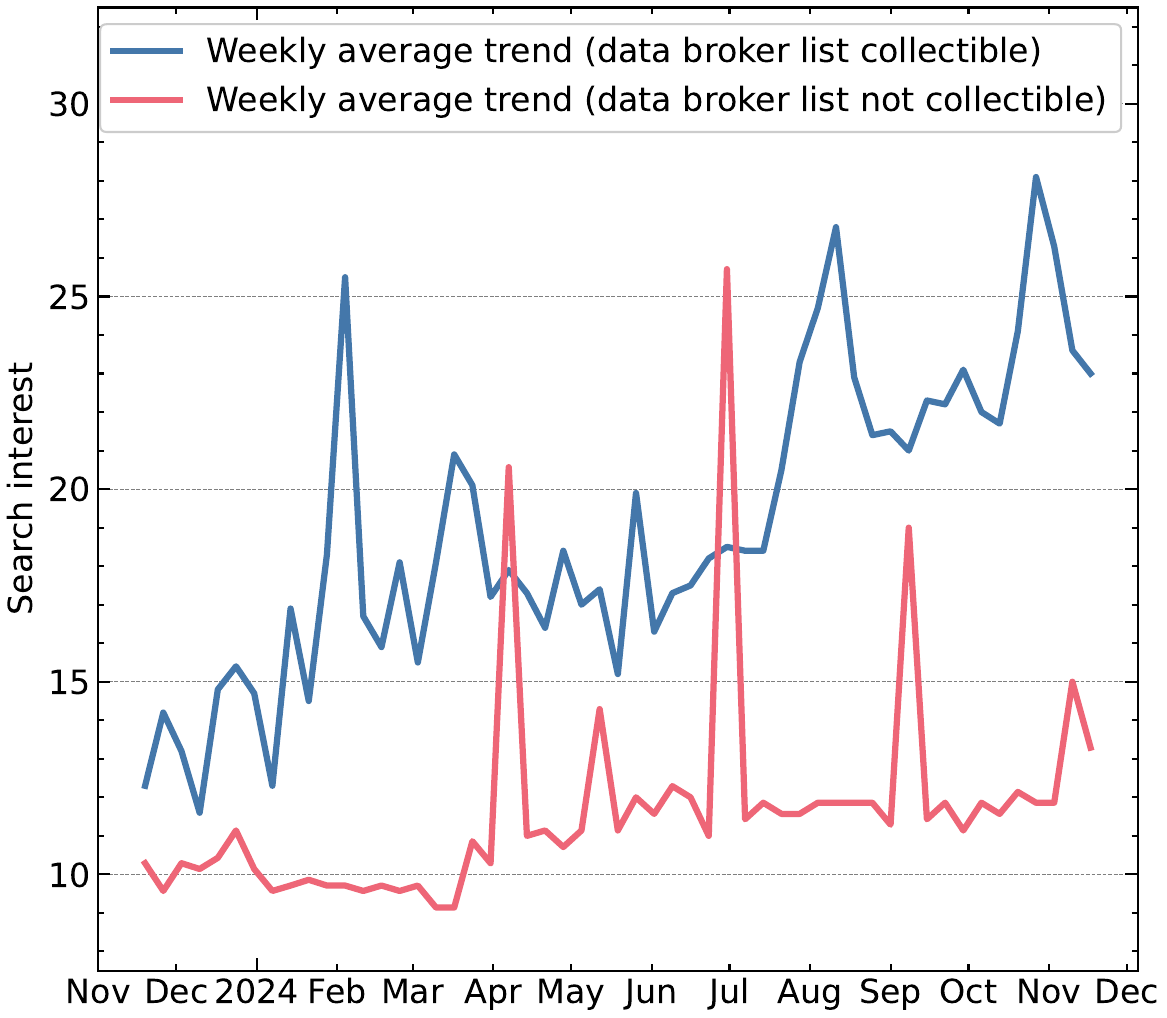}
     \caption{Weekly average of Google Trends search interest results for PII removal services' name from Nov 2023 to Dec 2024.}
     \label{fig:google_trend}
\end{figure}

\pb{Removal Services Data Summary.}
As result, Table \ref{table:service_summary} summarizes the 18 PII removal services that we find. 
Among them, \texttt{PrivacyBee}, \texttt{Aura}, \texttt{HelloPrivacy}, \texttt{PurePrivacy}, \texttt{ReputationDefender} and \texttt{Safe Shepherd} do not disclose the list of data brokers they cover. 
\texttt{Removaly} and \texttt{Desear.me} have been acquired/closed and are unavailable.
Therefore, the analysis in this paper focuses on the remaining 10 services for which we can collect the data broker list.
It should be noted that \texttt{PrivacyBot} is a free and open source service created by UC Berkeley, and while data brokers coverage is collectible, the project has been out of support since September 2021.
To briefly confirm the popularity of these PII removal services, we use Google Trends. Figure \ref{fig:google_trend} shows the Google search interests for the PII removal services' names over the last year. 
The blue line shows the average trend of the 10 services for which the data broker list can be collected; the red line shows the average trend for the 8 services for which their covered data broker list cannot be collected. We see here that the 10 services have noticeably higher Google attention, giving us confidence in their selection. As a supplement, we split the Google search interest for each service separately, please refer to Figure \ref{fig:google_trend_split} in Appendix \ref{appendix:a3}.

\subsection{Data Broker List}
\label{subsec:data_collection3}
 
\pb{PII Removal Services.}
For each of the above 10 PII removal services, from \texttt{March 11, 2024} to \texttt{November 18, 2024}, we use a script to automatically crawl the data broker coverage list on the official website of the PII removal service at the same time every week. This low-frequency scraping does not put a strain on the the PII removal service servers.
Note, in the data broker list published by \texttt{Optery}, there are 266 entries that only have names and no associated domains (\eg only PeopleSearcher, not peoplesearcher.com).
Therefore, we manually add the domains via Google search. As a result, these 10 PII removal services cover a total of 1,759 unique data brokers.

\pb{Government Registration.}
To enhance users' control over their PII, four states in the United States —-- namely Vermont \cite{vermont}, Texas \cite{texas}, California \cite{california} and Oregon \cite{oregon} —-- have implemented mandatory data broker registration. Registered data brokers are mandated to clearly outline the methods by which users can opt-out of their PII from these brokers.
Therefore, as a supplement, we also collect the lists of all registered data brokers that are publicly available from these four states. There are 528 (California), 481 (Vermont), 218 (Texas) and 191 (Oregon) data brokers, respectively. As a result, these four government registration sets contain a total of 764 unique data brokers. These additional data brokers are treated as part of our overall data broker dataset, allowing a more comprehensive understanding of the data broker landscape. It is important to note that the government-operated data broker registration websites do not provide opt-out services, so we are not regard them as PII removal services.

\begin{table*}[]
\begin{tabular}{lrrrrrr}
\hline
\textbf{\begin{tabular}[c]{@{}l@{}}PII removal\\ Services\end{tabular}} & \textbf{\begin{tabular}[c]{@{}r@{}}Discovery\\ Methods\end{tabular}} & \textbf{\begin{tabular}[c]{@{}r@{}}Data Broker\\ Collectible\end{tabular}} & \textbf{\begin{tabular}[c]{@{}r@{}}\# Covered\\ Data Brokers\end{tabular}} & \textbf{\begin{tabular}[c]{@{}r@{}}Subscription Price\\ (Monthly / Annual)\end{tabular}} & \textbf{Free Search} & \textbf{Free Removal} \\ \hline
DeleteMe \cite{DeleteMe}                                                   & Seed                                                                 & Y                                                                          & 759                                                                        & - / \$129                                                                                & N                    & N                     \\
Optery \cite{Optery}                                                      & Google Search                                                        & Y                                                                          & 125/243/743                                                                & \begin{tabular}[c]{@{}r@{}}(\$3.99, \$14.99, \$24.99) / \\ (\$39, \$149, \$249)\end{tabular}   & Y                    & N                     \\
PrivacyBot \cite{PrivacyBot}                                               & Google Search                                                        & Y                                                                          & 420                                                                        & Free                                                                                     & Y                    & Y                     \\
Kanary \cite{Kanary}                                                     & Google Search                                                        & Y                                                                          & 317                                                                        & \$16.99 / \$179.88                                                                         & Y                    & N                     \\
PrivacyDuck \cite{PrivacyDuck}                                               & Google Search                                                        & Y                                                                          & 274                                                                        & - / \$299.99                                                                             & N                    & N                     \\
Onerep \cite{Onerep}                                                   & Google Search                                                        & Y                                                                          & 233                                                                        & \$14.95 / \$99.96                                                                          & Y                    & N                     \\
Incogni \cite{Incogni}                                                  & Seed                                                                 & Y                                                                          & 217                                                                        & \$14.98 / \$89.88                                                                          & N                    & N                     \\
Mozilla Monitor \cite{MozillaBroker}                                       & Google Search                                                        & Y                                                                          & 196                                                                        & \$13.99 / \$107.88                                                                         & Y                    & N                     \\
EasyOptOuts \cite{EasyOptOuts}                                             & Google Search                                                        & Y                                                                          & 180                                                                        & - / \$19.99                                                                              & N                    & N                     \\
DataSeal \cite{DataSeal}                                               & Google Search                                                        & Y                                                                          & 115                                                                        & \$12.99 / \$99.99                                                                          & Y                    & N                     \\ \hline
PrivacyBee \cite{PrivacyBee}                                            & Google Search                                                        & N                                                                          & 885+                                                                       & - / \$197                                                                                & Y                    & N                     \\
PurePrivacy \cite{PurePrivacy}                                            & Google Search                                                        & N                                                                          & 200+                                                                       & \$9.99 / \$69.99                                                                           & Y                    & N                     \\
Aura \cite{Aura}                                                          & Seed                                                                 & N                                                                          & 20+                                                                        & \$15 / \$144                                                                               & N                    & N                     \\
HelloPrivacy \cite{HelloPrivacy}                                          & Google Search                                                        & N                                                                          & Unknown                                                                    & \$9.99 / \$99                                                                              & Y                    & N                     \\
ReputationDefender \cite{ReputationDefender}                                & Reddit                                                               & N                                                                          & Unknown                                                                    & - / -                                                                                    & N                    & N                     \\
Safe Shepherd \cite{SafeShepherd}                                       & Reddit                                                               & N                                                                          & Unknown                                                                    & \$13.95 / \$99.95                                                                          & N                    & N                     \\
Removaly                                                                & Google Search                                                        & N                                                                          & Unknown                                                                    & - / -                                                                                    & -                    & N                     \\
Deseat.me                                                               & Reddit                                                               & N                                                                          & Unknown                                                                    & - / -                                                                                    & -                    & N                     \\ \hline
\end{tabular}
\caption{Summary of 18 PII removal services, with Optery having different subscription levels.}
\label{table:service_summary}
\end{table*}

\subsection{Merging Data Brokers by eTLD+1}
\label{subsec:data_collection4}

We observe that occasionally the same data brokers appear under different domain names, \eg \url{people.yellowpages.com} and \url{yellowpages.com} refer to the same broker but are listed as separate domains in different PII removal services.
To merge these instances, we use Effective Top-Level Domain+1 (eTLD+1) to determine if they are the same website. \cite{mcquistin2024first}.
Overall, we find a total of 55 groups of data broker domains with same eTLD+1, as detailed in \ref{table:public_suffix} in Appendix \ref{appendix:a4}. 
For simplicity, we consider brokers with the same eTLD+1 as a single data broker and use the shorter domain name for representation, \eg we use \url{yellowpages.com} to replace \url{people.yellowpages.com}.

There are 6 PII removal services and 2 government registration data broker sets that have multiple domains with the same eTLD+1. These are \texttt{DeleteMe} (45 groups), \texttt{Vermont} (16), \texttt{PrivacyDuck} (6), \texttt{Kanary} (6), \texttt{Mozilla Monitor} (4), \texttt{Onerep} (4), \texttt{Optery} (2) and \texttt{Texas} (2).
After merging data brokers with the same eTLD+1, the ``\# Covered Data Broker'' column in Table \ref{table:service_summary} reflects the number of unique brokers for each service. This process results in a total of 2,024 unique data brokers (1,759 from 10 removal services and 265 from government registration), and we will only discuss these unique data brokers with different eTLD+1 later in this paper.

\begin{figure*}
     \centering
     \subfloat[][]{\includegraphics[width=.33\linewidth]{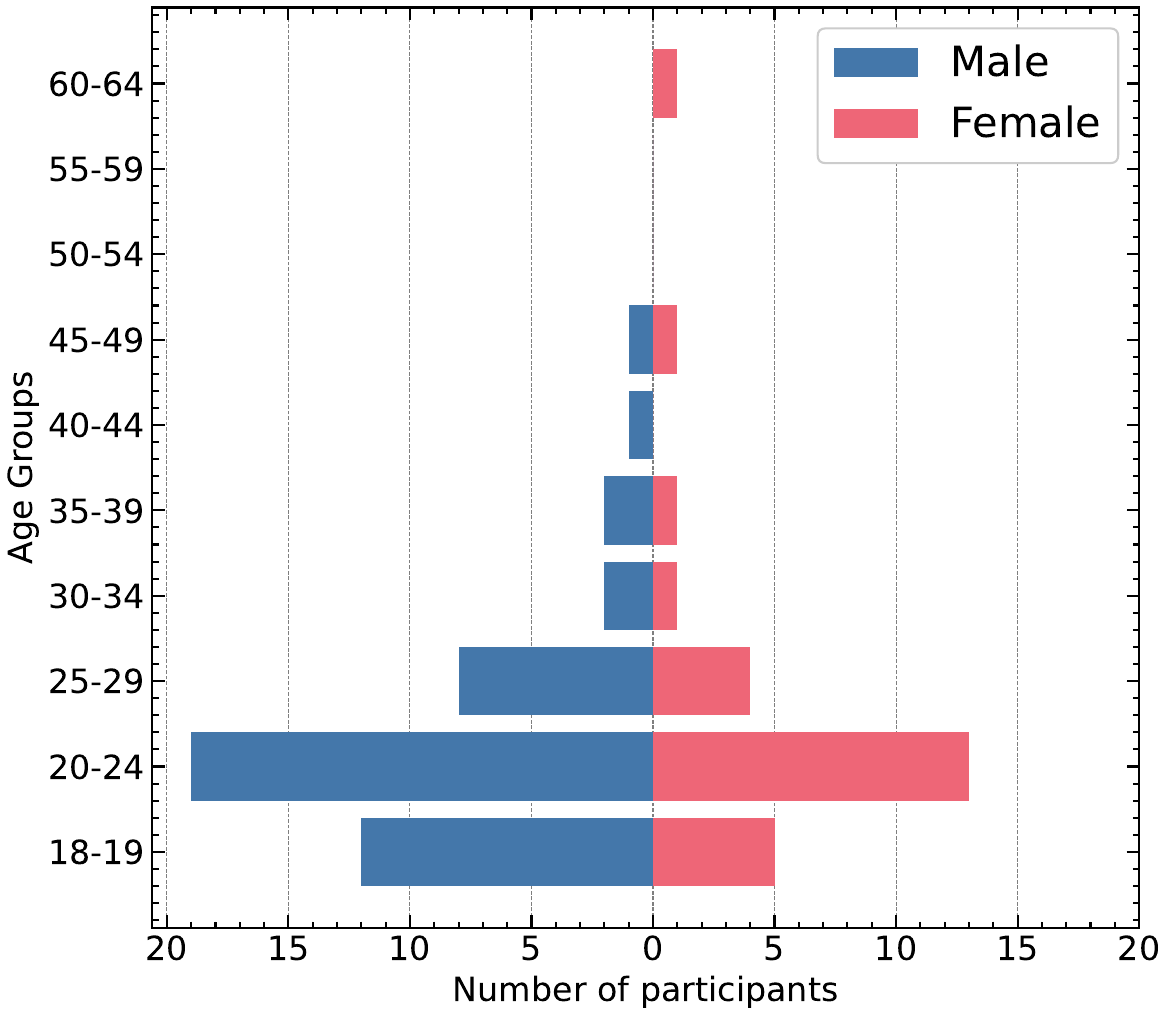}\label{fig:age_gender}}
     \subfloat[][]{\includegraphics[width=.33\linewidth]{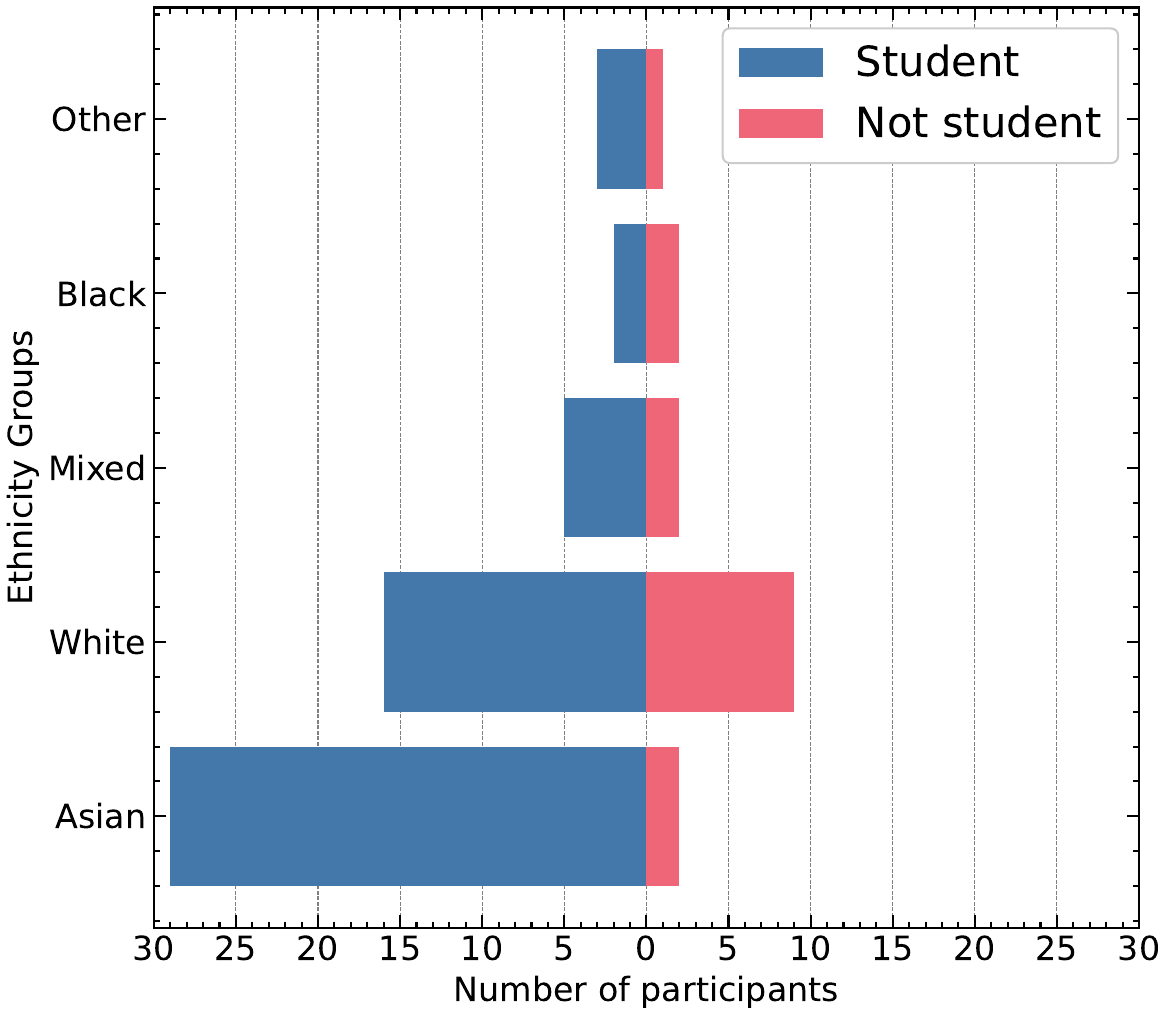}\label{fig:ethnicity_student}}
     \subfloat[][]{\includegraphics[width=.33\linewidth]{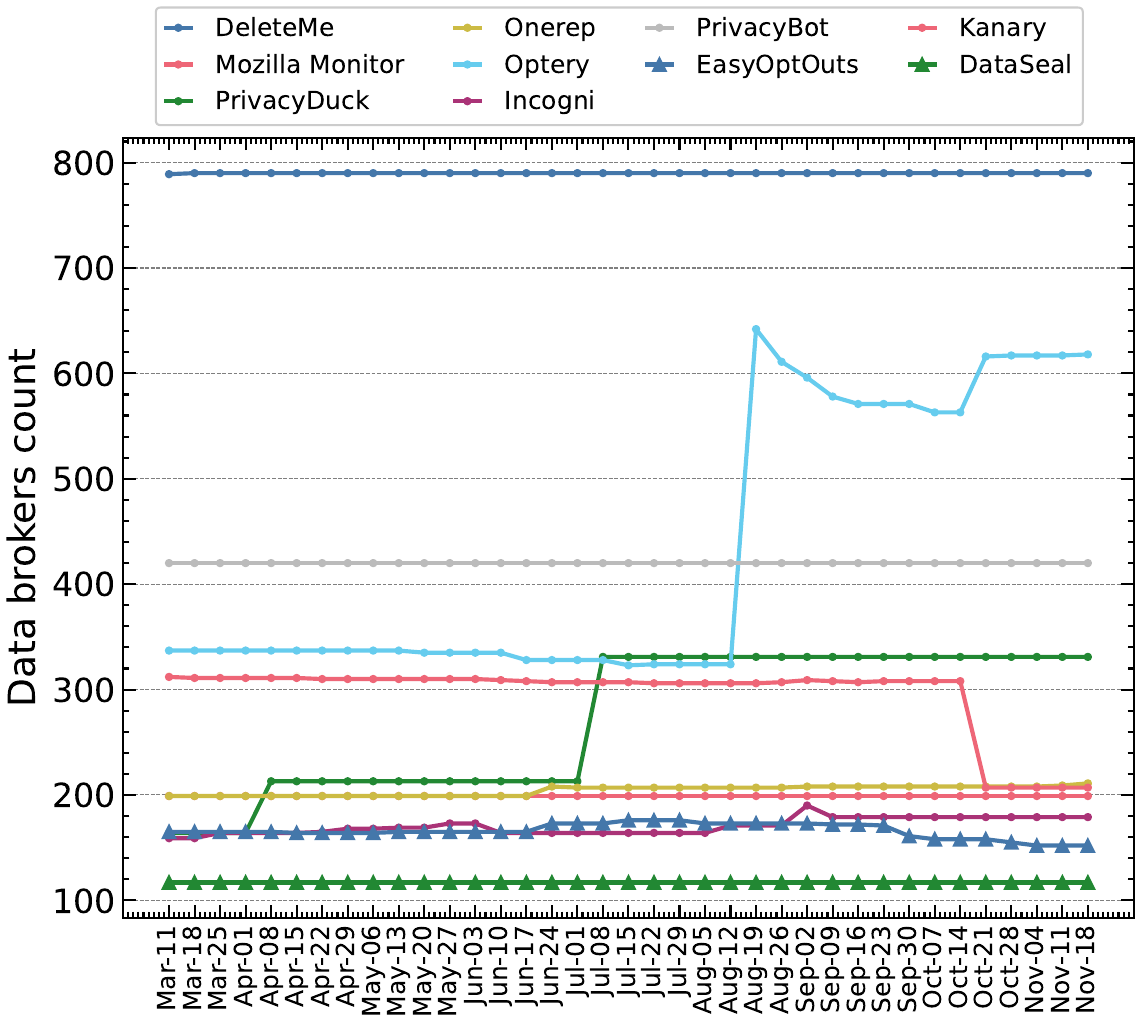}\label{fig:data_broker_weekly_distribution}}
     \caption{(a) Demographic information of 71 recruited participants: age and gender. (b) Demographic information of 71 recruited participants: ethnicity and student status. (c) Weekly distribution of data broker coverage for 10 PII removal services (during data collection period)}
\end{figure*}

\subsection{User Study Methodology}
\label{subsec:data_collection5}
To evaluate the efficacy of the PII removal services, we further conduct a user study, recruiting participants to use the services.

\pb{Service Selection.}
We first select the services to evaluate. 
Due to funding constraints that limit how many subscriptions we can pay for, we must balance the desire for covering all services vs.\ the desire to get a large number of samples for each service.
To begin, we only consider services that offer monthly subscriptions, leaving a total of 6 options (see ``Subscription Price'' column in Table \ref{table:service_summary}). Among these, \texttt{Onerep} and \texttt{Mozilla Monitor} are in partnership and have a high overlap in their data broker coverage. Therefore, we choose the cheaper option (\texttt{Mozilla Monitor}) to maximize our ability to recruit more users.
Additionally, we exclude \texttt{DataSeal} due to its limited data broker coverage.
We also exclude the free service \texttt{PrivacyBot}, as it is deprecated and requires users to manually configure Google OAuth credentials. This complexity likely poses challenges for recruited users, since many lack technical backgrounds.
As a result, we select the remaining 4 services to evaluate their efficacy: \texttt{Optery}, \texttt{Kanary}, \texttt{Mozilla Monitor}, and \texttt{Incogni}.
Note, \texttt{Optery} offers three different subscription plans, which differ only in their data broker coverage. To eliminate the influence of subscription price on the service's performance, we select a plan (\$14.99) that is most comparable to that of the other services. This choice provides \texttt{Optery} with a data broker coverage of 243. 
Thus, in our user study, these 4 services claim to cover removals from 659 unique data brokers.

\pb{Participant Recruitment.}
We recruit a total of 80 participants (20 per removal service). As 9 participants withdrew from the study before completion, we ultimately receive valid data from 71 participants, with 18 from Prolific and 53 from Northwestern University.
Prolific is an online research platform, and provides the recruitment and management of participants for online research. With Prolific, we can easily communicate with the participants, check the progress of experiment and paying users, while Northwestern University is able to increase the reliability and scale of the results without excessive cost. By merging these two recruitment strategies, we enhance the diversity of the sample.

We use Prolific's demographic data to gather information on Prolific participants, and ask participants from Northwestern University to complete the same questionnaire to collect demographic information for that group. Note, we require participants to be currently residing in the United States, as the data broker industry is primarily located there. 
We leave analysis of wider geographical trends to future work. 
For context, Figure \ref{fig:age_gender} and \ref{fig:ethnicity_student} shows the age-gender and status-ethnicity relationship of all 71 participants (all participants are over 18 years of age). Overall, the ratio of male to female participants is 63.4\% to 36.6\%, with students making up 77.5\% of the total. 

We note that our sample size (71) is relatively modest and, therefore, caution should be exercised when generalizing the findings to a broader population, specifically in terms of demographics, technical expertise, and geographical factors. 
That said, we believe our study offers important insights into the effectiveness of PII removal services and lays a solid groundwork for future research with larger samples.

\pb{Experimental Setup.}
For each participant, we first provide registration and subscription instructions across the different removal services. For Prolific users, we cover their subscription fees through Prolific. However, they need to follow our instructions to pay the subscription fee on their own. For Northwestern university users, we provide offline guidance to participants. We assist them in completing the registration on their computers and cover the subscription fee for them using a temporary credit card.
During the registration process, participants are required to input the requested personal information to help the service retrieve records about them in the data broker database (refer to Table \ref{table:required_pii} in \S\ref{subsec:service_characterization2} for details).
After 30 days of subscription,\footnote{In line with the claims made by these four PII removal services, we consider a 30-day period sufficient. 
The four providers assert that they can achieve significant progress within 10 days to a few weeks. We confirm this assumption in \S\ref{subsec:crowdsourcing_experiment3}.} we ask participants to share their service removal progress data with us, using an in-house browser plugin.\footnote{Plugin available for use by researchers at \url{https://github.com/xxx/xxx} (anonymous in the review stage)}
The browser plugin enables us to automatically confirm that the participant has successfully registered for the service, as well as verify the validity of the participant's sending of the service removal progress data (see Appendix \ref{appendix:a5} for more information).

Finally, to better understand the accuracy of the records retrieved by the PII removal service, we invite participants to self-assess the accuracy of the records retrieved by the service. Participants are asked to review all the records retrieved by their service and categorize each record into one of the following three categories: 
\one \textbf{Correct}, this record contains correct information about the participant; 
\two \textbf{Incorrect}, this record does not contain information about the participant, it is about someone else; 
and
\three \textbf{Unsure}, it is unclear, there is not enough information to determine, or unable to open the data broker web page. 
At the end of the experiment, we instruct all participants to cancel their subscription (providing the simple instructions).

\subsection{Ethical Considerations}
\label{subsec:data_collection6}

For the user study, we inform participants about the detailed procedures of the experiment upfront, and require each participant to sign a consent form (please refer to Appendix \ref{appendix:a6} for the consent form).
Information about the PII removal services and data brokers is available on their websites, and we encourage participants to learn about the PII removal service in detail before participating in the experiment.
We inform participants that they can withdraw from the experiment at any time. Participants are rewarded \$40 upon completion of the experiment, and participants who provide the service retrieval record accuracy assessment receive an additional \$15. 
As well as the payment, participants further benefit from the free removal of their PII as we cover the subscription cost for the removal service.

For participants' personal information, we collect only the name and email of the user (from Northwestern University) for contact and payment purposes based on the user's consent, and delete their information after the experiment is completed. Beyond that, we do not collect any PII from participants in the experiment. The PII removal service results that participants send to us do \emph{not} contain any of the participant's personal data, nor do they contain any information about what data was specifically removed from the data broker by the PII removal service.
For the results sent to us by the participants, we use them only to assess the efficacy of the PII removal service and the accuracy of the retrieved records. All data is stored securely and only the authors can access them.
All detailed procedures for the experiment are approved by the authors' home institution and we have obtained IRB approval.\footnote{Protocol code is HSP-2024-0023}

%% file: Section/4.Services_Characterization.tex
\section{Overview of PII Removal Services and Data Brokers (RQ1)}
\label{sec:service_characterization}

In this section we explore the characteristics of the 10 PII removal services and the 1,759 data brokers they cover. Note, this excludes the 265 unique data brokers that were found in the the government registration, but not covered by any removal services. This includes the data broker coverage for each service, the amount of PII a user needs to provide when subscribing, alongside the industry categories that the data brokers belong to.

\begin{figure*}
     \centering
     \subfloat[][]{\includegraphics[width=.4\linewidth]{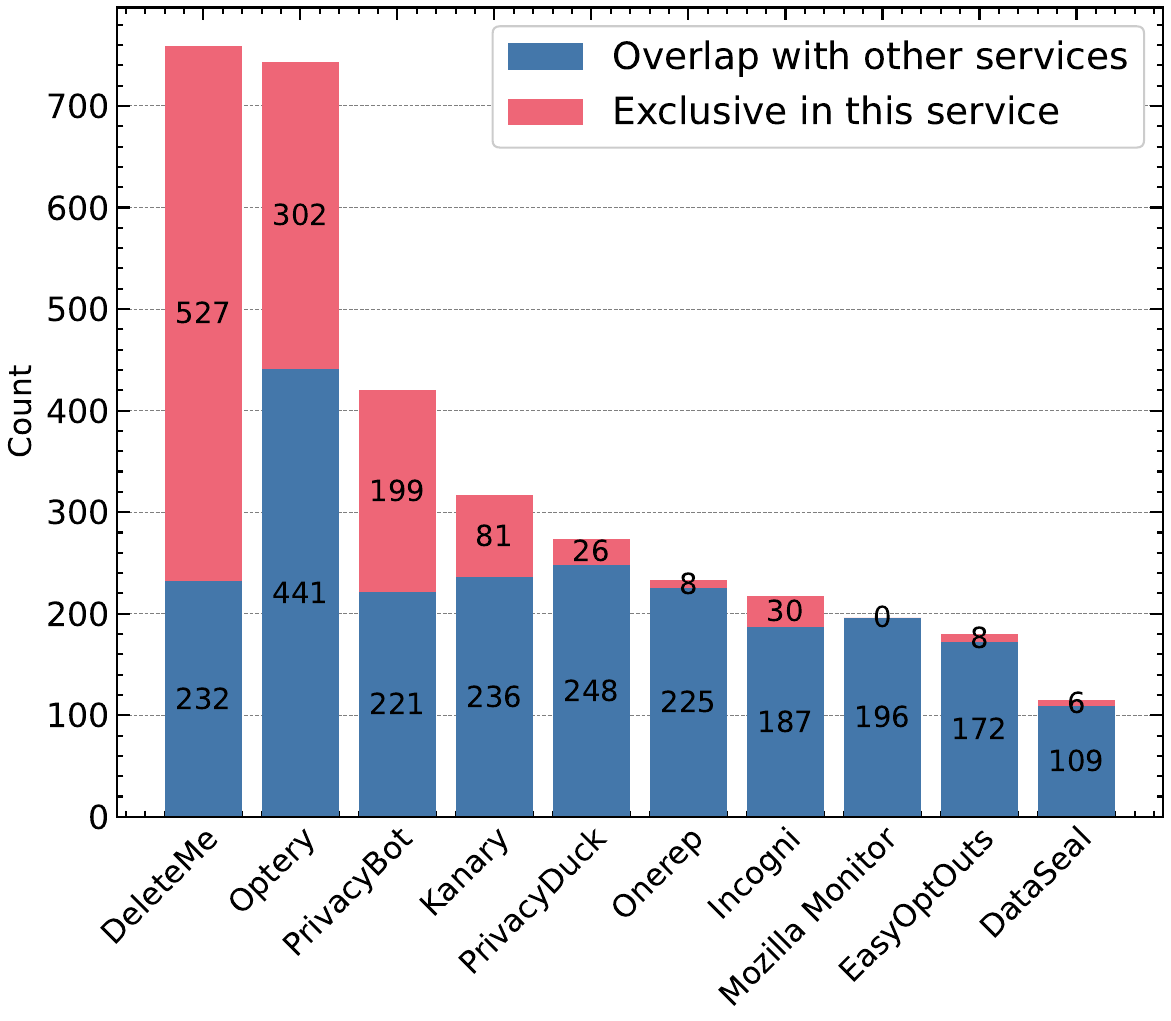}\label{fig:data_broker_overlap_exclusive}}
     \subfloat[][]{\includegraphics[width=.4\linewidth]{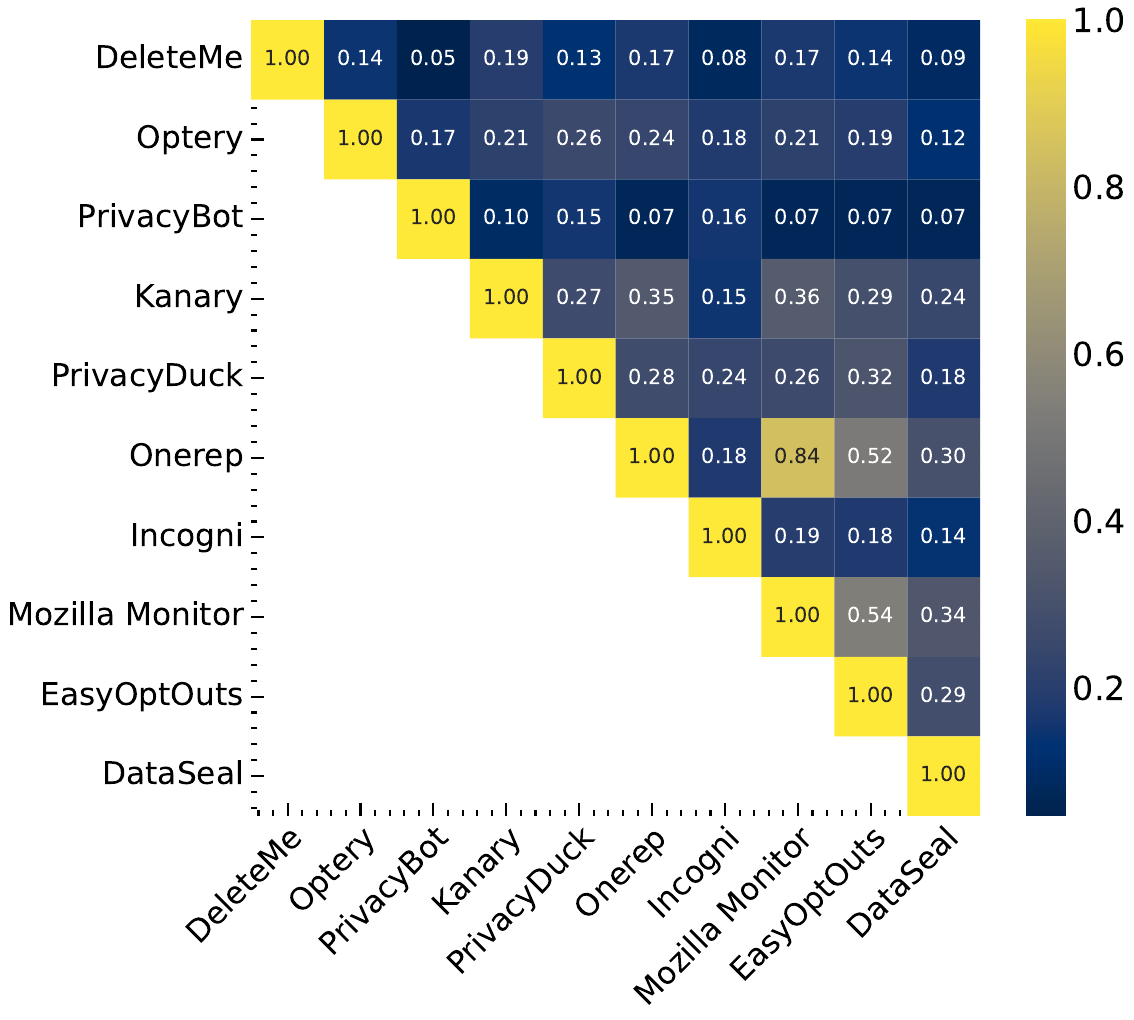}\label{fig:data_broker_jaccard_similarity}}
     \caption{(a) Overlap and exclusivity of per-service data broker coverage. (b) Heatmap of Jaccard similarity between PII removal services.}
\end{figure*}

\begin{figure*}
     \centering
     \subfloat[][]{\includegraphics[width=.4\linewidth]{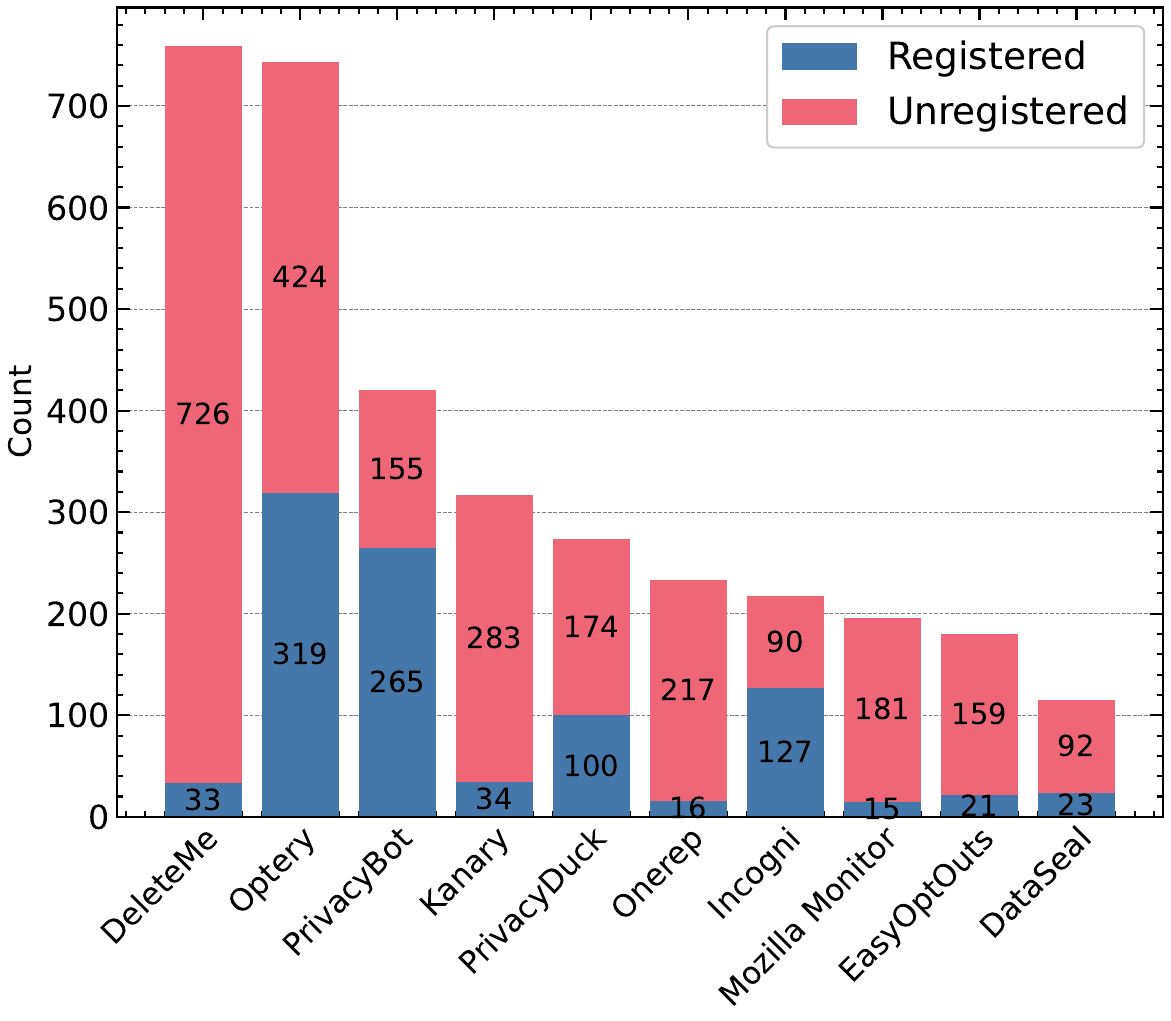}\label{fig:data_broker_registered_unregistered}}
     \subfloat[][]{\includegraphics[width=.4\linewidth]{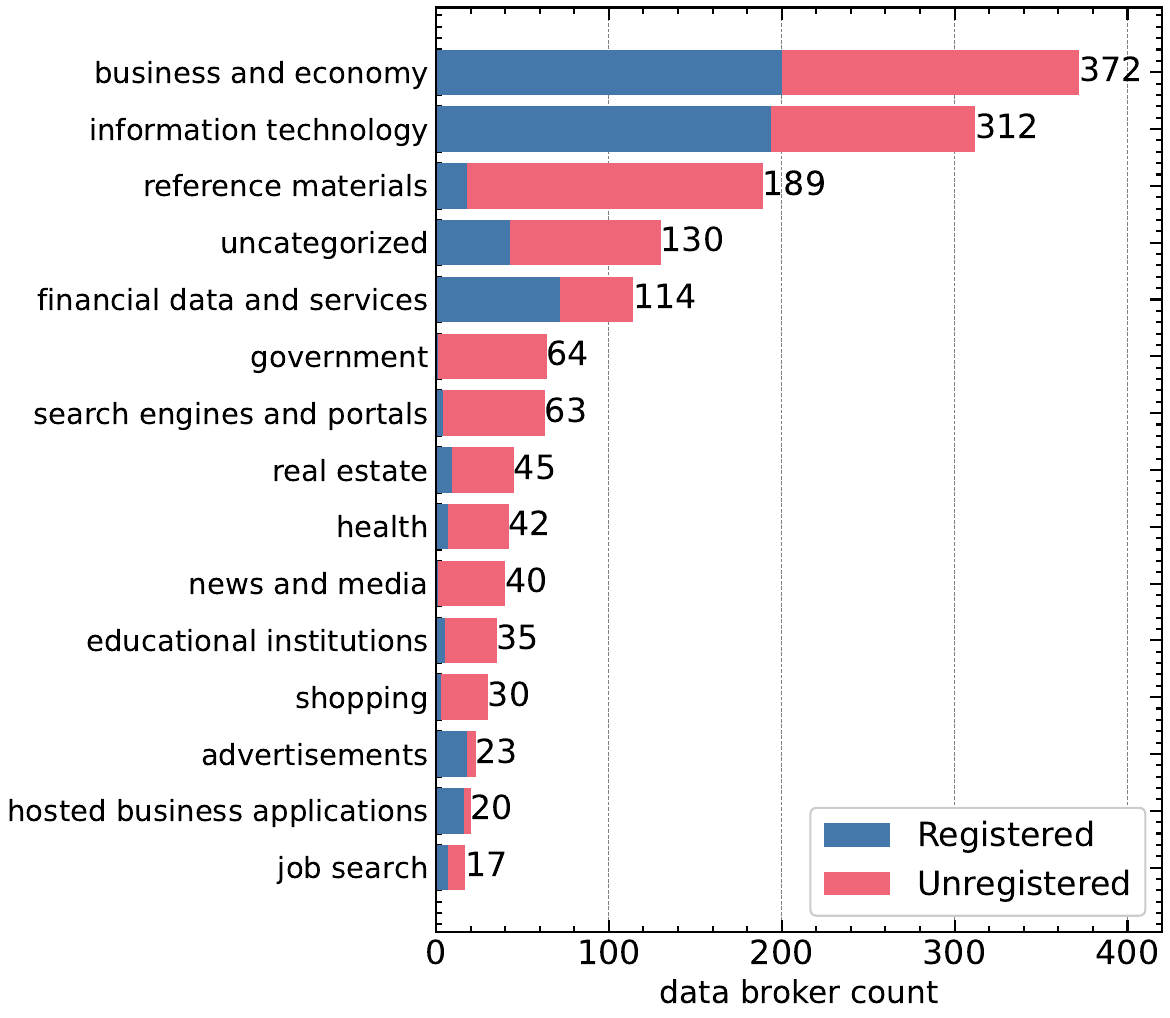}\label{fig:category}}
     \caption{(a) Registered and unregistered of per-service data broker coverage. (b) The top 15 industry categories to which the data broker belongs to, and the number of registered (blue) and unregistered (red) in each category.}
\end{figure*}

\begin{table*}[]
\resizebox{2\columnwidth}{!}{%
\begin{tabular}{lll}
\hline
\textbf{Service} & \textbf{Mandatory PII}                                          & \textbf{Optional PII}                         \\ \hline
Optery &
  \textbf{Full Name}, \textbf{Email}, \textbf{City Address}, Date of Birth &
  Physical Address, Phone Number, Family \& Relatives, Employment, Gender, ID, LinkedIn URL \\
DeleteMe &
  \textbf{Full Name}, \textbf{Email}, \textbf{Physical Address}, Date of Birth, Phone Number &
  Family \& Relatives, Employment, Gender, ID \\
Kanary &
  \textbf{Full Name}, \textbf{Email}, \textbf{City Address}, Year of Birth &
  Date of Birth, Physical Address, Phone Number, Employment \\
DataSeal         & \textbf{Full Name}, \textbf{Email}, \textbf{City Address}, Year of Birth                   & Phone Number, Family \& Relatives             \\
EasyOptOuts &
  \textbf{Full Name}, \textbf{Email}, \textbf{Physical Address}, Year of Birth, Phone Number, Family \& Relatives &
  - \\
PrivacyDuck      & \textbf{Full Name}, \textbf{Email}, \textbf{Physical Address}, Date of Birth, Phone Number & -                                             \\
Onerep           & \textbf{Full Name}, \textbf{Email}, \textbf{City Address}                                  & Physical Address, Date of Birth, Phone Number \\
Incogni          & \textbf{Full Name}, \textbf{Email}, \textbf{Physical Address}                              & Date of Birth, Phone Number                   \\
PrivacyBot       & \textbf{Full Name}, \textbf{Email}, \textbf{Physical Address}, Date of Birth, Phone Number & -                                             \\
Mozilla Monitor  & \textbf{Full Name}, \textbf{Email}, \textbf{City Address}, Date of Birth                   & -                                             \\ \hline
\end{tabular}%
}
\caption{Mandatory and optional PII for each PII removal services, with common mandatory PII in bold. The city address is a location specific to the city, and the physical address is a more detailed address that usually includes the street name, house number and ZIP code.}
\label{table:required_pii}
\end{table*}

\subsection{Removal Services' Data Broker Coverage}
\label{subsec:service_characterization1}

We first examine the data broker coverage of the PII removal services. Arguably, services that support information removal from more data brokers would be more effective in protecting users' PII. 

\pb{Overall Coverage.}
The ``\# Covered Data Brokers'' column in Table \ref{table:service_summary} shows the number of data brokers (after removing the same eTLD+1) covered by each removal service. \texttt{DeleteMe} and \texttt{Optery} are the two services with the largest data broker coverage, 759 and 743 respectively.
However, with the growth of the data broker industry, the data broker coverage of the PII removal service is evolving. 
Figure \ref{fig:data_broker_weekly_distribution} shows the weekly distribution of the number of data brokers covered per service across the data collection period. Over this period, the data broker coverage of these 10 PII removal services increase by an average of 36.3. Among them, \texttt{Optery} increases the most, with 281 new data brokers added to its list. 
However, surprisingly, \texttt{EasyOptOuts} and \texttt{Kanary} actually decrease their data broker coverage, by 13 and 105, respectively. This may be due to a change in the data broker's opt-out method that causes the service to no longer support the broker.

Given that there are approximately 5,000 data brokers in operation \cite{maximize, knowledgesourcing}, the coverage offered by the removal services (1,759 data brokers, see \S\ref{subsec:data_collection3}) is arguably insufficient. 
This indicates that most of the data brokers in-the-wild remain difficult for users to remove their PII from. However, we argue that the data brokers covered by these PII removal services represent a fairly comprehensive subset of those that do support opt-outs.

\pb{Data Broker Coverage Overlap.}
We observe that certain data brokers appear frequently across multiple PII removal services. 
To understand this, we examine whether each removal service's data broker coverage is exclusive, or overlapping with others. 
Figure \ref{fig:data_broker_overlap_exclusive} shows the number of overlapping (blue) and exclusive (red) data brokers for different services. Compared to other services, \texttt{DeleteMe} and \texttt{Optery} have a larger number of exclusive data brokers, accounting for 69.7\% and 40.6\% of their list, respectively. In contrast, \texttt{Onerep}, \texttt{EasyOptOuts}, and \texttt{DataSeal} each have fewer than 10 exclusive data brokers, while \texttt{Mozilla Monitor} has none. This suggests that \texttt{DeleteMe} and \texttt{Optery} access more unique data brokers that others do not, potentially enhancing their effectiveness in protecting user privacy and attracting more users.

To further analyze the overlap, we calculate the Jaccard index, shown in Figure \ref{fig:data_broker_jaccard_similarity} as a heatmap. The Jaccard index quantifies the degree of overlap between two sets, focusing only on shared elements and ignoring sequential and duplicate elements, making it well suited for set-based comparisons.
This index ranges from 0 (no common elements) to 1 (identical sets). The overall similarity between services is low (average 0.21), except for \texttt{Mozilla Monitor} and \texttt{Onerep}, which have a similarity of 0.84. 
This is because \texttt{Mozilla Monitor} partners with \texttt{Onerep}, leading to nearly identical broker lists.
Despite this overlap, the 10 PII removal services collectively cover 1,759 different data brokers (excluding data brokers from government registration, see \S\ref{subsec:data_collection3}), with only 10 brokers appearing across all services. 
This indicates a degree of uniqueness among services, suggesting they may target different user groups. This also means that users may benefit from subscribing to multiple removal services to remove PII from a wider range of data brokers.

\pb{Registered and Unregistered Data Brokers.}
Recall that four states in the United States require data brokers to register in a public listing.
They currently list 764 unique registered data brokers (see \S\ref{subsec:data_collection3}). 
However, there are actually about 5,000 data brokers in operation today \cite{maximize, knowledgesourcing}, and only about 15\% are registered, which shows that most data brokers in-the-wild still lack proper supervision.
We therefore further examine the number of registered and unregistered data brokers covered by each removal service.

The results are shown in Figure \ref{fig:data_broker_registered_unregistered}.
Overall, 71.7\% of data brokers across the 10 PII removal services are \emph{not} registered in any of the government databases. \texttt{DeleteMe}, which has the largest coverage of data brokers, covers only 4.3\% registered brokers. In contrast, \texttt{Optery} covers the highest number of registered data brokers, accounting for 42.9\%.
The low registration rate suggests that data brokers are not being adequately regulated, which undoubtedly has a direct impact on the transparency of personal data removals. Additionally, there is significant room for improvement in the coverage of data brokers by the services: Incorporating registered data brokers could expand the service's coverage.

\subsection{Required PII by Removal Services}
\label{subsec:service_characterization2}
The PII removal services are designed to assist subscribed users in automatically removing PII from data brokers. Thus, upon subscription, users are required to input some of their private information (\eg name, email, date of birth), which helps the service check the corresponding records from the data broker databases. Therefore, we next examine the types of user private information required by different services.

Table \ref{table:required_pii} shows the mandatory and optional PII that users can provide to each removal service. Overall, the PII required varies for each service, and \textbf{Full Name}, \textbf{Email} and \textbf{City Address} are three common PII that the user must provide to each service. Perhaps unsurprisingly, services that cover more data brokers, also request more PII.
For example, \texttt{Optery} and \texttt{DeleteMe} require 10 and 9 items of PII, respectively (they also have the most data broker coverage). This is likely because different data brokers require different types of PII to retrieve records. This also suggests that the amount of PII a user provides may affect the accuracy of the retrieved records (we will evaluate in a later section \S \ref{subsec:crowdsourcing_experiment2}).

We note that this observation may mean that removal services, themselves, become a privacy risk in the case of breaches or data resale.
This is not beyond the realms of possibility, \eg \texttt{Onerep}'s CEO has been exposed as having ties to multiple personal search sites \cite{onerep1, onerep2}.

\subsection{Data Broker Categories}
\label{subsec:data_broker_characterization3}

We next examine the industry categories to which data brokers belong. To do this, we use multiple website categorization engines to label the data broker domains. In order to validate the results, we randomly sample 50 data broker domains to manually check the categories are sensible (see Appendix \ref{appendix:a7} for detailed results). We find that \emph{Forcepoint ThreatSeeker} can identify 98\% of the correct categories. Thus, we use \emph{Forcepoint ThreatSeeker} as the categorization engine here.

Figure \ref{fig:category} shows the distribution of the top 15 data broker industry categories. Overall, the data broker industry is quite diverse. Unsurprisingly, the largest proportions are found in \textit{business and economy}, \textit{information technology}, and \textit{reference materials}, accounting for 19.0\%, 16.0\%, and 9.7\%, respectively. Additionally, it also encompasses areas such as \textit{government}, \textit{health}, \textit{shopping}, and \textit{job search}.
The wide variety of data brokers indicates that they have permeated many sectors. These websites not only host a significant amount of PII, but may also contain sensitive data related to personal relationships, shopping preferences, and health conditions.

We further examine the number of government registered and unregistered data brokers in the different categories, also shown in Figure \ref{fig:category}. Recall, these registrations are legally mandated in four US states.
Data brokers that are classified as related to the economy and finance have a high number of registrations (\eg 53.8\% in \textit{business and economy}, 63.1\% in \textit{financial data and services}). Other than that, the number of brokers in other categories is significantly lower. One potential explanation is that data brokers in the finance categories are more strictly regulated, leading to higher rates of registration. However, this does leave many data broker categories which have worryingly low registration rates.

Overall, the concentration of data brokers in sectors like \textit{business and economy} (19.0\%) and \textit{information technology} (16.0\%) is hardly surprising given the economic value and digital nature of personal data today. However, the proliferation across a wide array of other categories, including sensitive areas such as \textit{health} (2.14\%) and \textit{job search} (0.86\%), underscores the reach of data brokers into nearly all aspects of individuals' lives . This widespread presence, coupled with the finding that a substantial majority (71.7\%) of these entities operate without being registered with government authorities, paints a concerning picture. This lack of oversight means that many data brokers handling sensitive personal information may be operating with limited accountability, making it challenging for individuals to understand how their data is being used.

\takehomebox{\textbf{Take homes}: 
\one PII removal services cover different data brokers, suggesting that users need to use multiple removal services to get full coverage. The overlap in data broker coverage between services is low, with an average Jaccard similarity of just 0.21. 
\two 71.7\% of data brokers are not registered with the government authorities, highlighting the current lack of regulation. 
\three Removal services also collect PII, asking users to submit at a minimum their Full Name, Email and City Address. Removal services with larger data broker coverage tend to require additional PII from subscribers. 
\four Data brokers are distributed across various industries, with business (19.0\%) and information technology (16.0\%) being prominent.}

%% file: Section/6.Crowdsourcing.tex
\begin{figure*}
     \centering
     \subfloat[][]{\includegraphics[width=.4\linewidth]{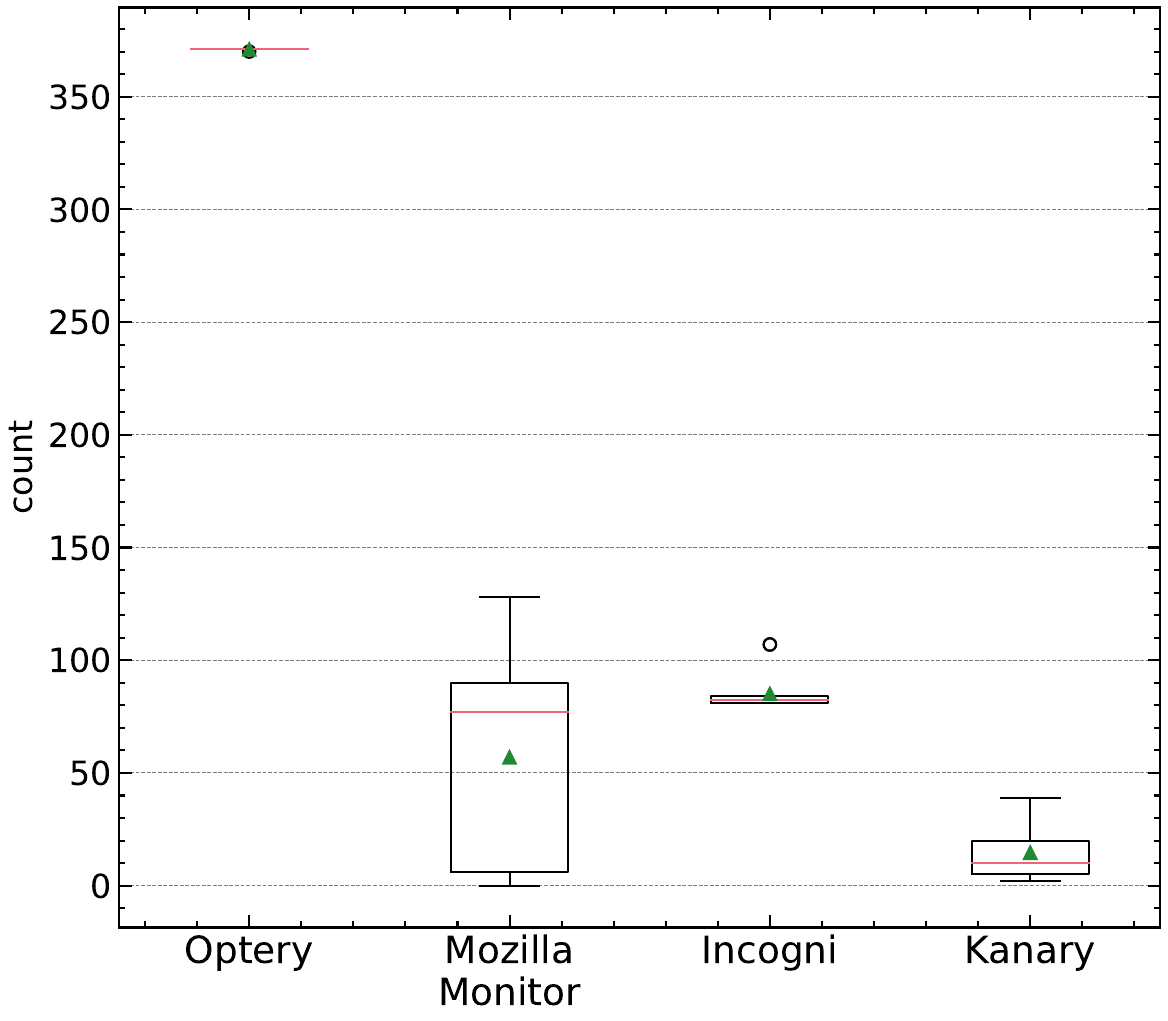}\label{fig:exposed_brokers}}
     \subfloat[][]{\includegraphics[width=.4\linewidth]{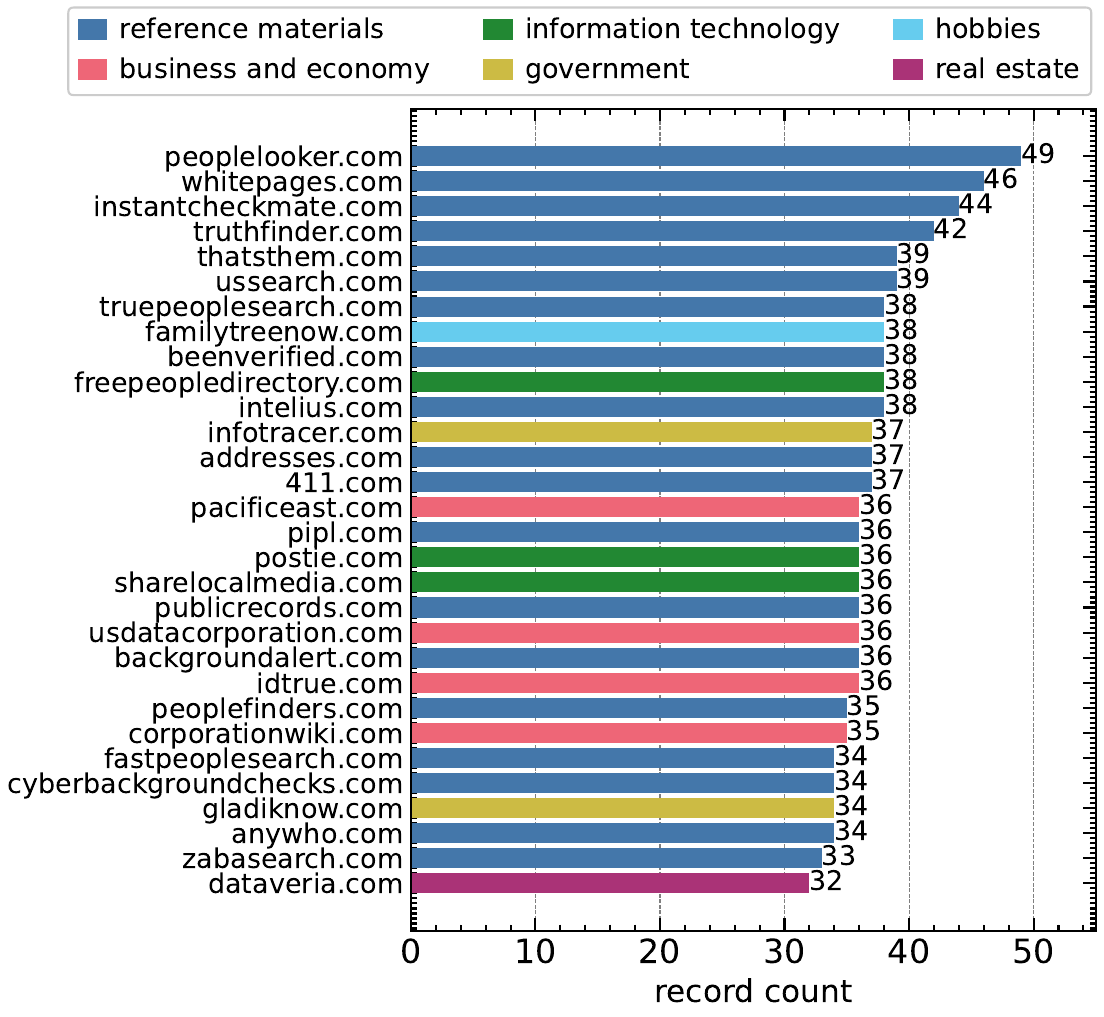}\label{fig:top30_broker_category}}
     \caption{(a) Boxplot of the number of records retrieved by each PII removal service for its users. (b) The top 30 data brokers that appeared most frequently in the retrieved records, and their industry categories.}
\end{figure*}

\section{PII Identified by Removal Services (RQ2)}
\label{sec:crowdsourcing_experiment}

To assess the effectiveness of PII removal services, it is crucial to understand their ability to locate user records held by the data brokers. The number of records a removal service can identify, on behalf of its users, can therefore serve as a indicator of its operational reach.
In this section, we examine the number of PII records retrieved by each of the evaluated removal services for their users, as well as whether the record points to the correct person.

\subsection{Number of PII Records Identified Per Service}
\label{subsec:crowdsourcing_experiment1}

We start with examining the number of records retrieved by each PII removal service. We argue that this can help us better understand the retrieval capabilities of PII removal services in their own data broker coverage.

Figure \ref{fig:exposed_brokers} shows the number of records discovered to be stored in each of the PII removal services.
Note that \texttt{Optery} retrieves almost the same number of 
records for each participant, because they do not distinguish between ``PII not found'' (\ie the PII removal service does not find a record of the user on the data broker) and ``PII removed'' (\ie the PII removal service has removed the user's record from the data broker, so the user's record can no longer be found on the data broker).
Thus, each participant has the same number of records. 
In terms of the number of records identified, \texttt{Kanary} has the worst performance among the four services, even though it has largest data broker coverage (317).
It finds an average of 14.6 records per user, even though it claims to cover 317 brokers.
This indicates that a larger public data brokers coverage list does not necessarily guarantee that PII can then be retrieved and removed from these data brokers. 
We next examine the specific data broker domain retrieved by the service. We find that some data brokers appear more frequently, indicating that they gather more PII. 
Figure \ref{fig:top30_broker_category} shows the top 30 data brokers in terms of the number of records that are discovered on them. We color code the graph based on their industry categories (see \S\ref{subsec:data_broker_characterization3} for categorization details). 
In our previous observations, we found that \textit{business and economy} and \textit{information technology} represent the largest share of data brokers (see Figure \ref{fig:category}).
However, the \textit{reference materials} category reflects the largest source of successfully discovered user records.

One potential explanation is that data brokers in different industries may utilize distinct data sources, and an individual's online behavior can affect whether their information is captured by a specific data broker. For instance, a data broker in the \textit{business and economy} sector might concentrate on an individual's financial transaction records, tax payment records, and related information.
Whereas a data broker in the \textit{shopping} industry could focus on shopping histories, shipping addresses, and other purchasing-related data. The participants we recruit appear more frequently as data brokers for \textit{reference materials} (\eg people search site, yellow pages site \etc).

\begin{figure*}
     \centering
     \subfloat[][]{\includegraphics[width=.33\linewidth]{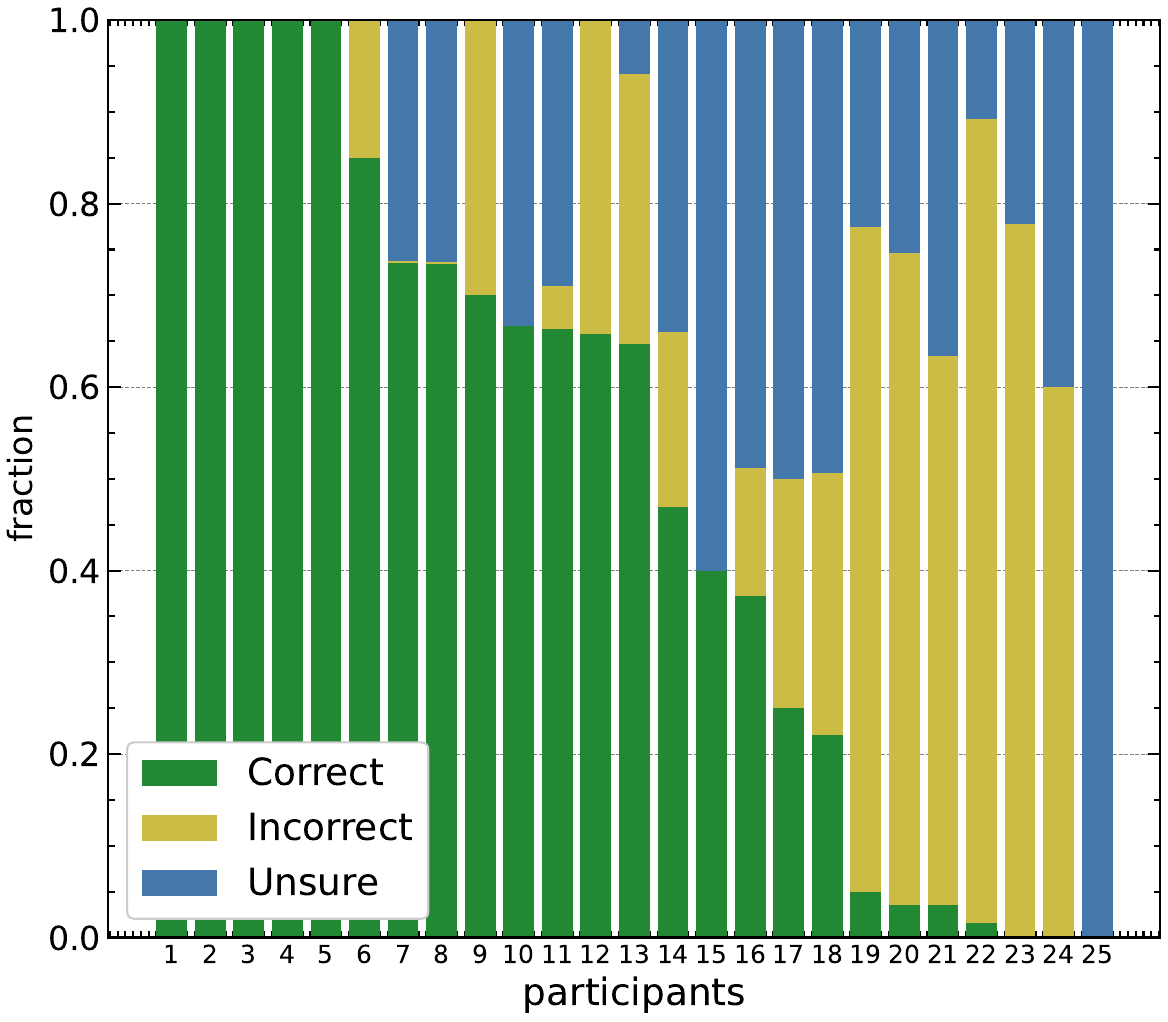}\label{fig:accuracy_result_prec}}
     \subfloat[][]{\includegraphics[width=.33\linewidth]{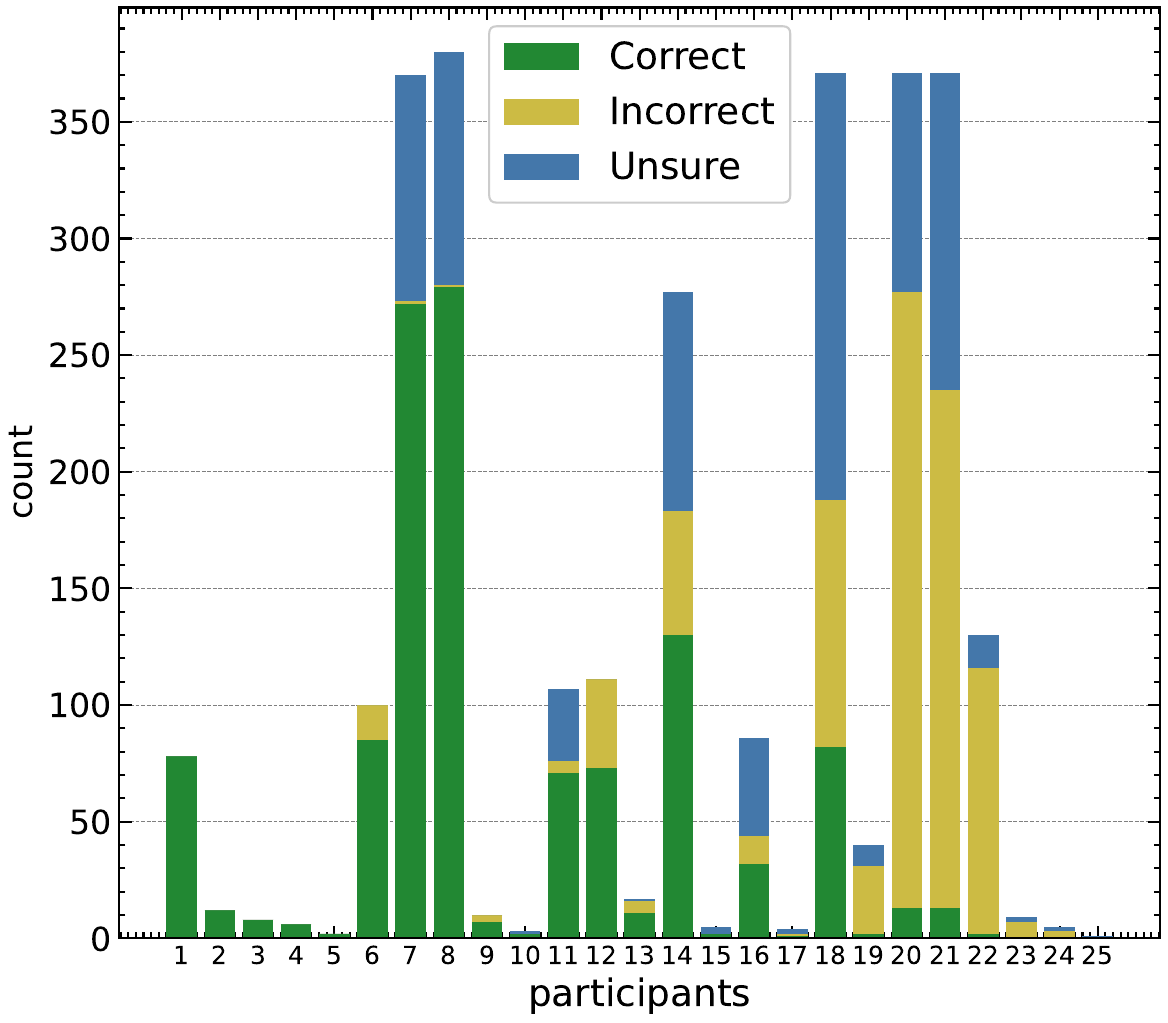}\label{fig:accuracy_result_count}}
     \subfloat[][]{\includegraphics[width=.33\linewidth]{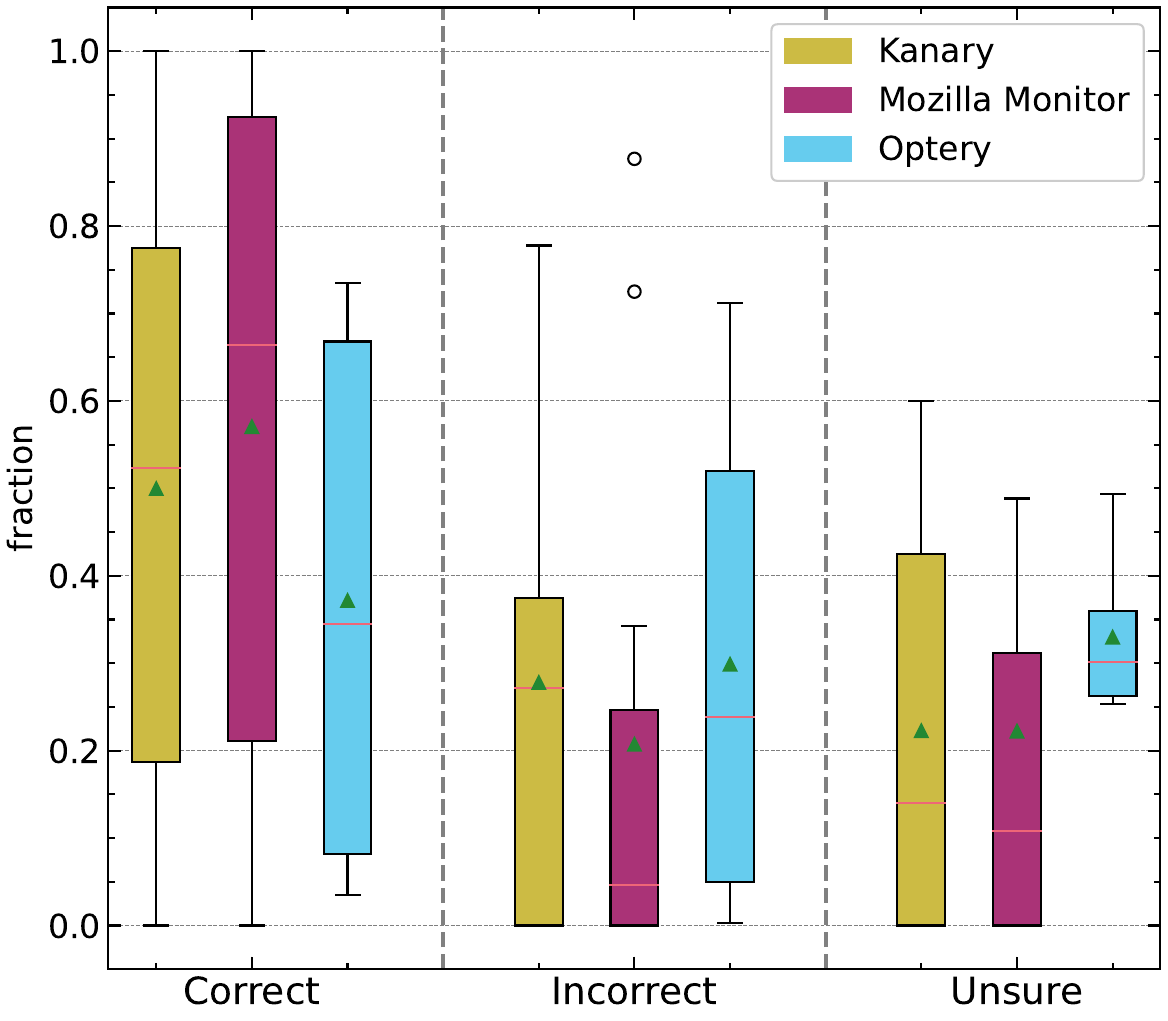}\label{fig:self_evaluation_service}}
     \caption{(a) Percentage distribution of accuracy of retrieved records per user. (b) Number distribution of accuracy of retrieved records per user (in the corresponding positions with (a)). (c) Boxplot of the distribution of accuracy of records retrieved by each PII removal service.}
\end{figure*}

\subsection{Accuracy of PII Removal Services}
\label{subsec:crowdsourcing_experiment2}

Recall that we invited participants to self-assess the accuracy of the records removed, and categorize all their retrieved records into three categories: \textbf{Correct}, \textbf{Incorrect}, and \textbf{Unsure} (see \S\ref{subsec:data_collection5} for more details).
Note, \texttt{Incogni} does not show users the specific PII contained in the retrieved records, so \texttt{Incogni} participants cannot self-assess their accuracy. Since some participants chose to withdraw from this part of the study, as a result, we collect accuracy self-assessment results from 25 participants.

Figure \ref{fig:accuracy_result_prec} presents the fraction of the three categories for each participant, and the Figure \ref{fig:accuracy_result_count} shows the actual counts of each category for each participants (in the corresponding positions with Figure \ref{fig:accuracy_result_prec}). 
Overall, the accuracy of the service's retrieval records is relatively low, with only 41.1\% of all records marked as correct. The percentage of incorrect and unsure were 30.7\% and 28.2\%, respectively.
We observe that participants who retrieved a smaller number of records tended to have higher accuracy. For example, for participants who retrieved fewer than 100 records, the removal service is 9.74\% more correct compared to those who retrieved more than 100 records.

We further examine the accuracy of the retrieved records on a per-service basis. 
Figure \ref{fig:self_evaluation_service} illustrates the boxplot for each category of the three services. 
Surprisingly, \texttt{Mozilla Monitor} has the least data broker coverage among the three services (see Table \ref{table:service_summary}), and has the least amount of PII required from users (see Table \ref{table:required_pii}).
Yet it achieves the highest retrieval accuracy. It has an average of 57.0\% correct records per user. 
We also observe that the more records a service retrieves does not necessarily lead to more correct records.
For example, \texttt{Optery} is significantly ahead of other services in the number of records it retrieves for its users, However, on average, 29.9\% of the records per user are incorrect and 33.0\% are unsure.

Overall, the PII removal service performs poorly in terms of retrieval record accuracy.
This may result in not removing the correct user records at the data broker. Instead, these incorrect records may belong to someone else with partially identical PII (\eg same name or same date of birth).
Consequently, this would \emph{not} reduce the risk of PII exposure for the paid subscription users, and lead users to wrongly believe their data has been removed. This is highly problematic and highlights the necessity for services to improve the accuracy of record identification, especially when users have limited PII to provide at the time of subscription.

\takehomebox{\textbf{Take homes}: 
\one The four PII removal services shows significant variation in record discovery performance.
\texttt{Kanary} performs the worst --- it covers 317 brokers, yet it only discovers an average of 14.6 records per user. 
\two Data brokers classified in the \textit{Reference materials} sector collect the most user PII during the experiment. 
\three The removal service's record retrieval accuracy is only 41.1\%. This indicates many removed records do not actually belong to the subscribe users. \texttt{Mozilla Monitor} has the highest accuracy (57.0\%) despite having the smallest data broker coverage, and requiring the least PII from user.}

\section{Efficacy of PII Record Removal (RQ3)}
\label{subsec:crowdsourcing_experiment3}

\begin{figure*}
     \centering
     \subfloat[][Optery]{\includegraphics[width=.25\linewidth]{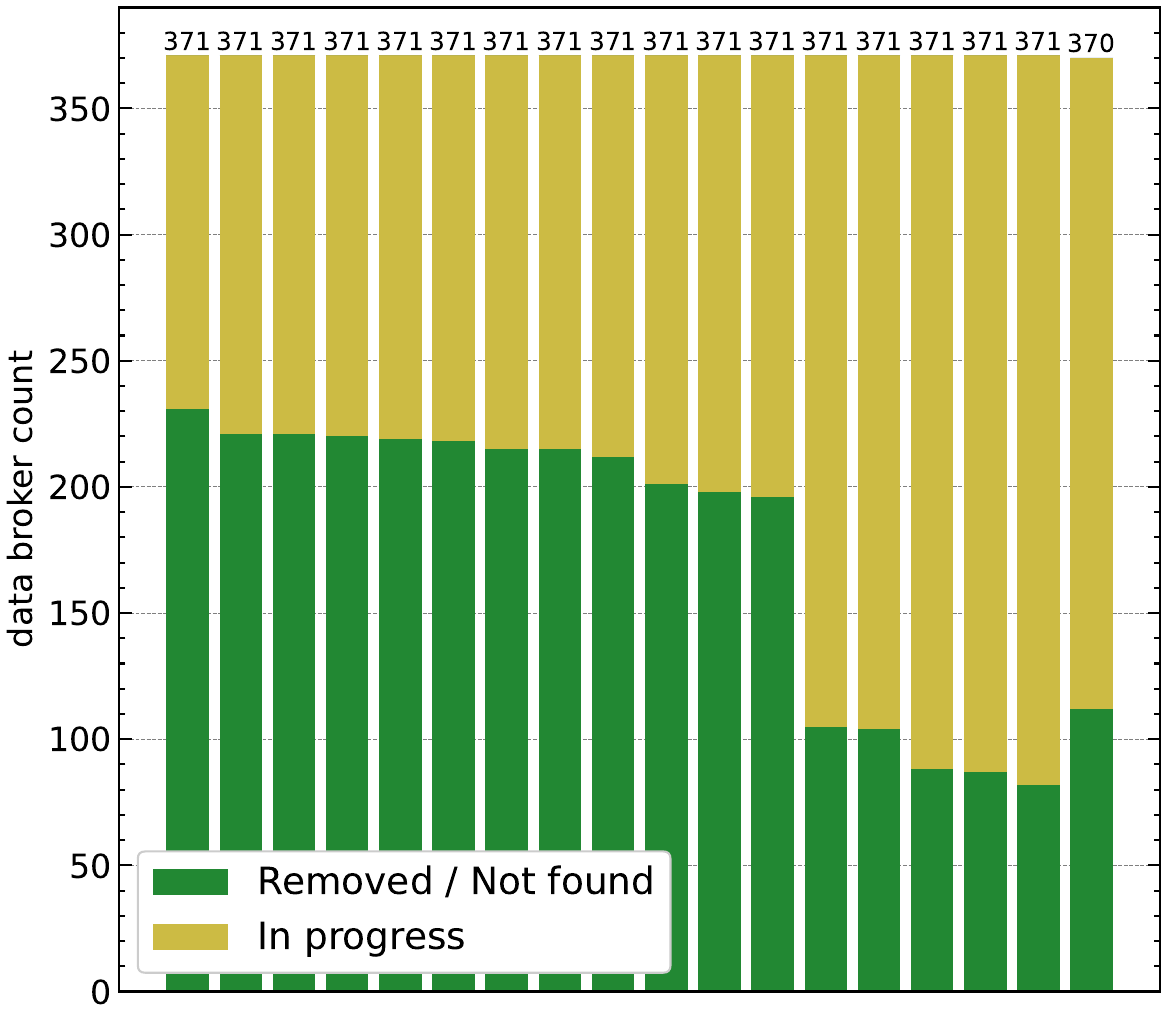}\label{fig:optery_result}}
     \subfloat[][Mozilla Monitor]{\includegraphics[width=.25\linewidth]{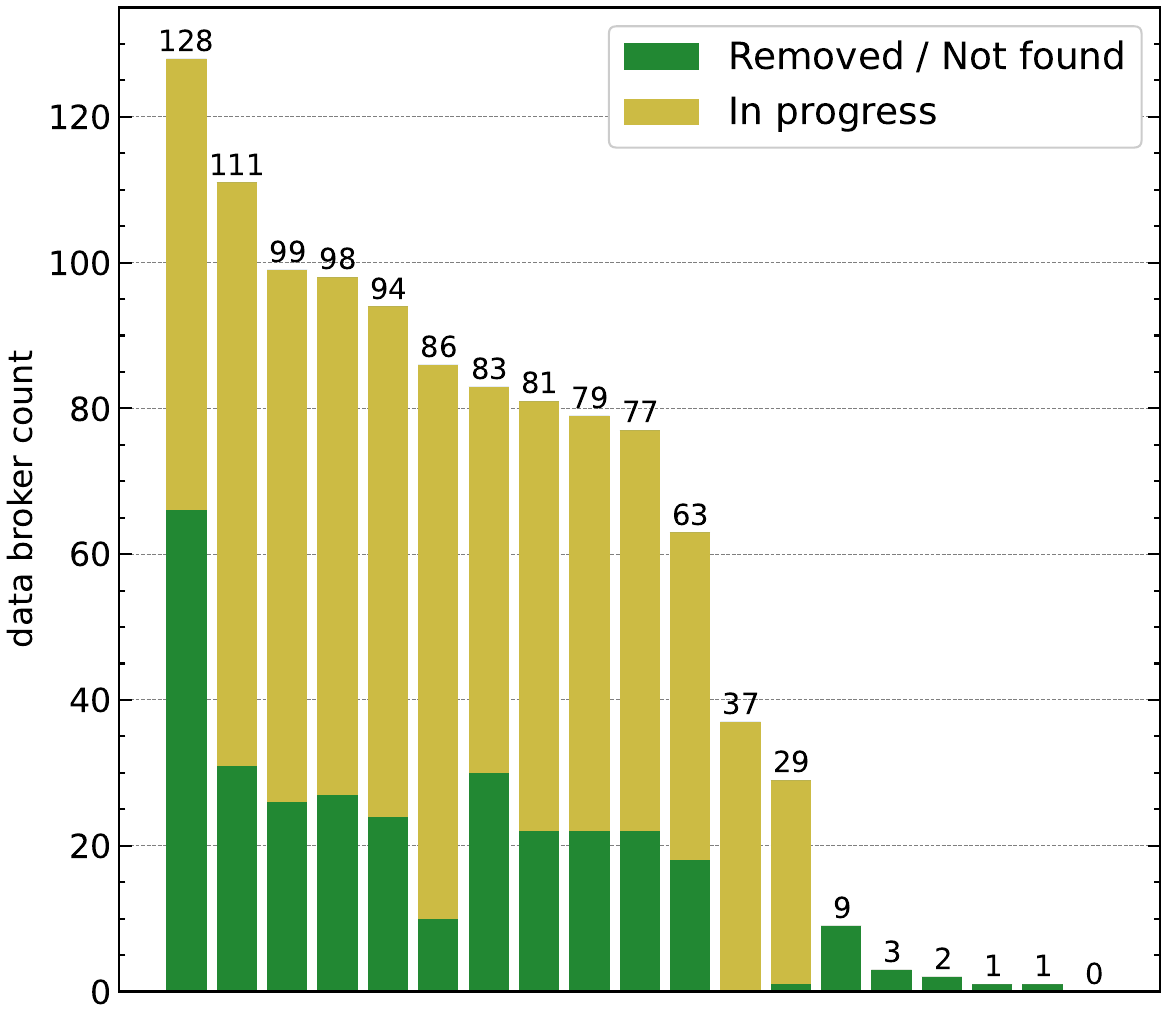}\label{fig:mozilla_result}}
     \subfloat[][Incogni]{\includegraphics[width=.25\linewidth]{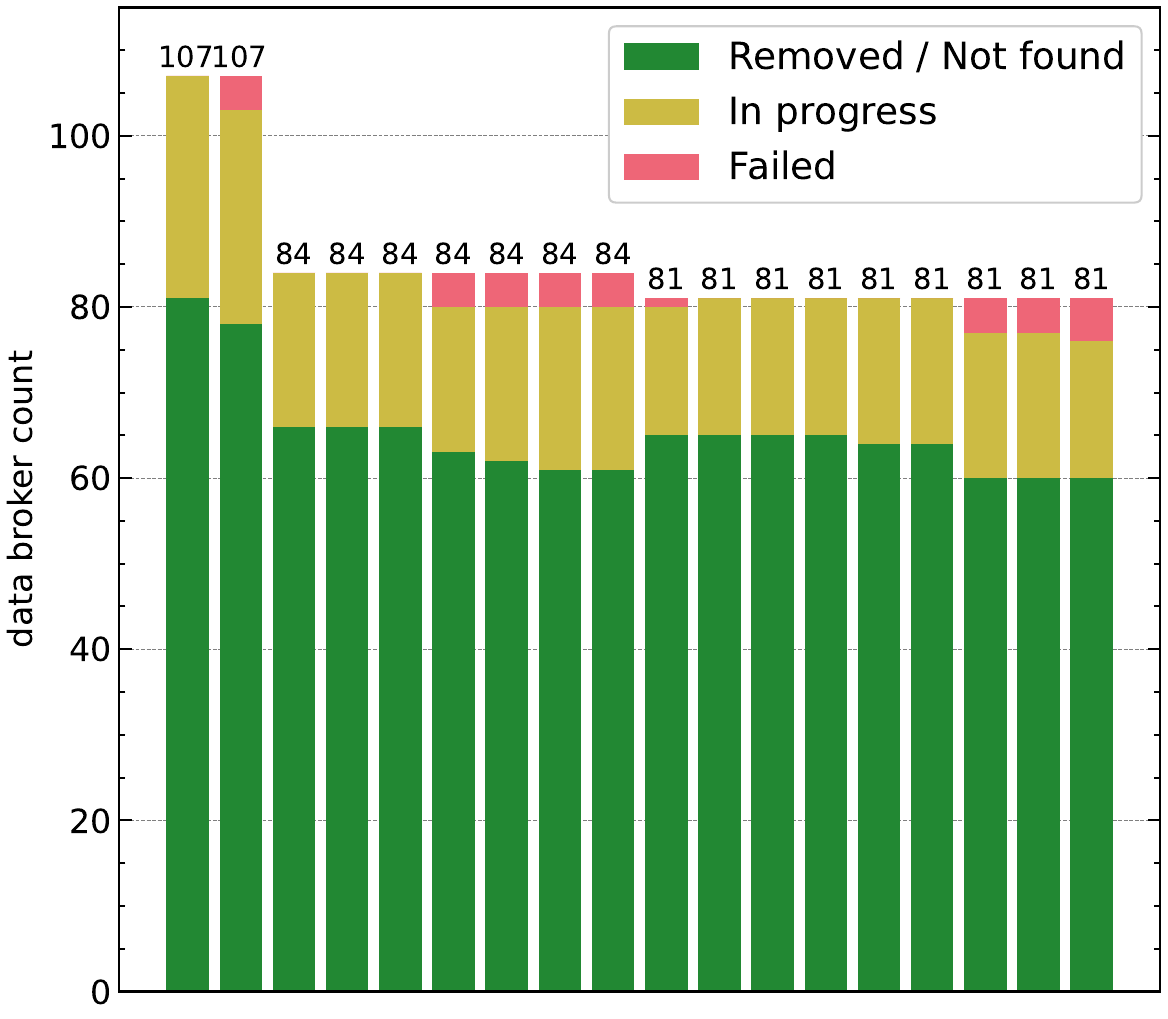}\label{fig:incogni_result}}
     \subfloat[][Kanary]{\includegraphics[width=.25\linewidth]{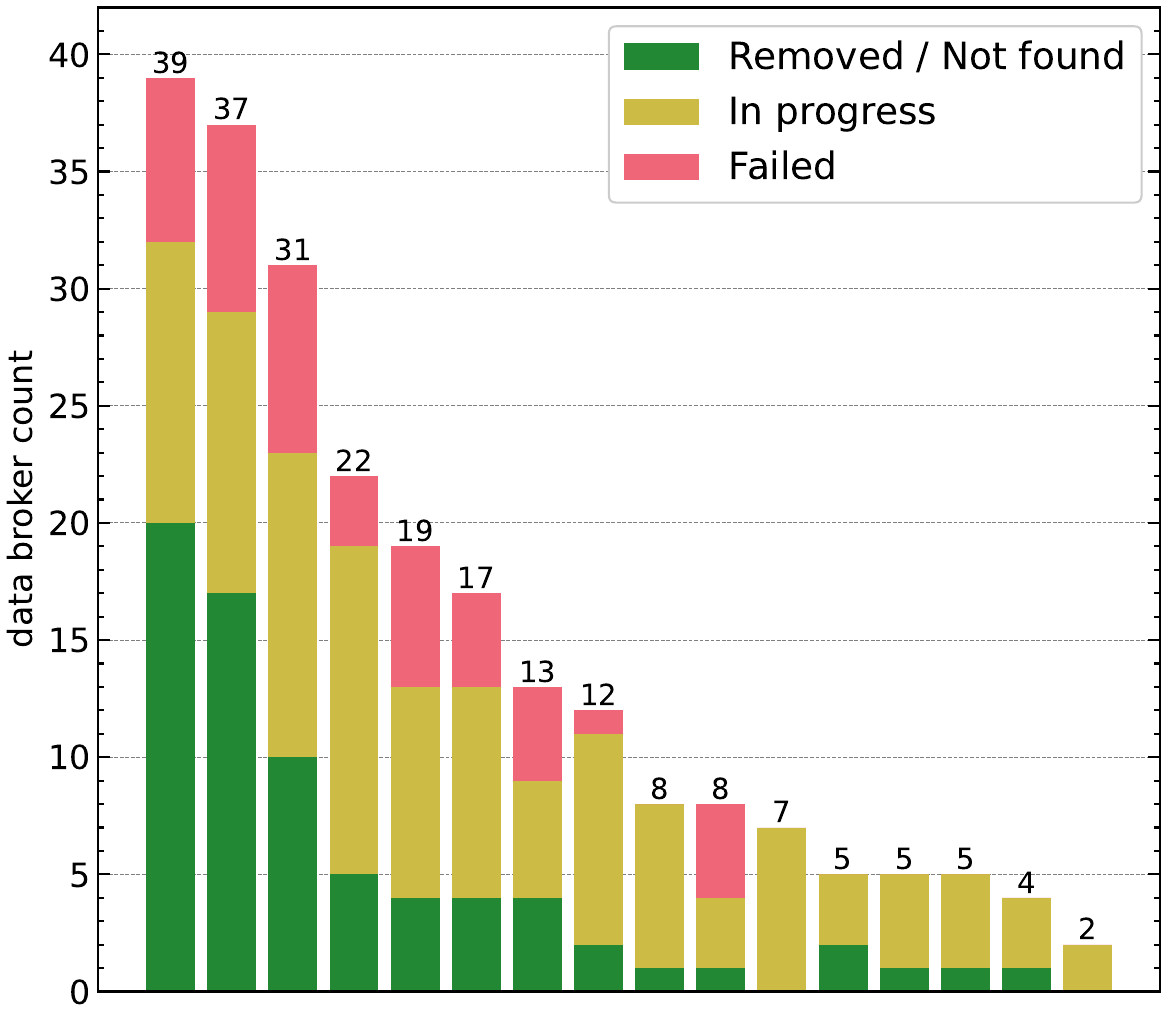}\label{fig:kanary_result}}
     \caption{The removal efficiency of PII removal service for its users within one month of subscription.}
     \label{fig:removal_result}
\end{figure*}

\begin{figure*}
     \centering
     \subfloat[][Optery]{\includegraphics[width=.25\linewidth]{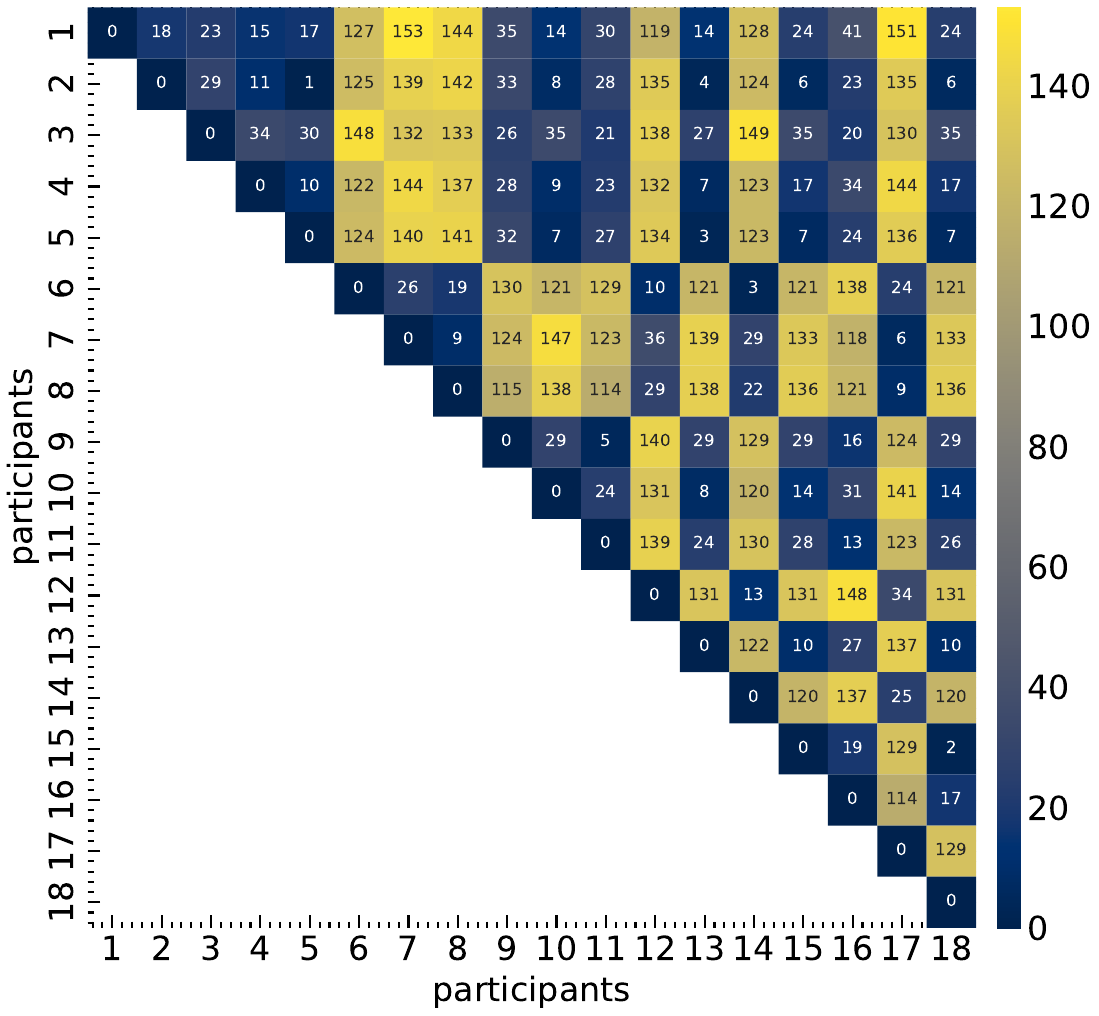}\label{fig:optery_remove_inprogress_overlap}}
     \subfloat[][Mozilla Monitor]{\includegraphics[width=.25\linewidth]{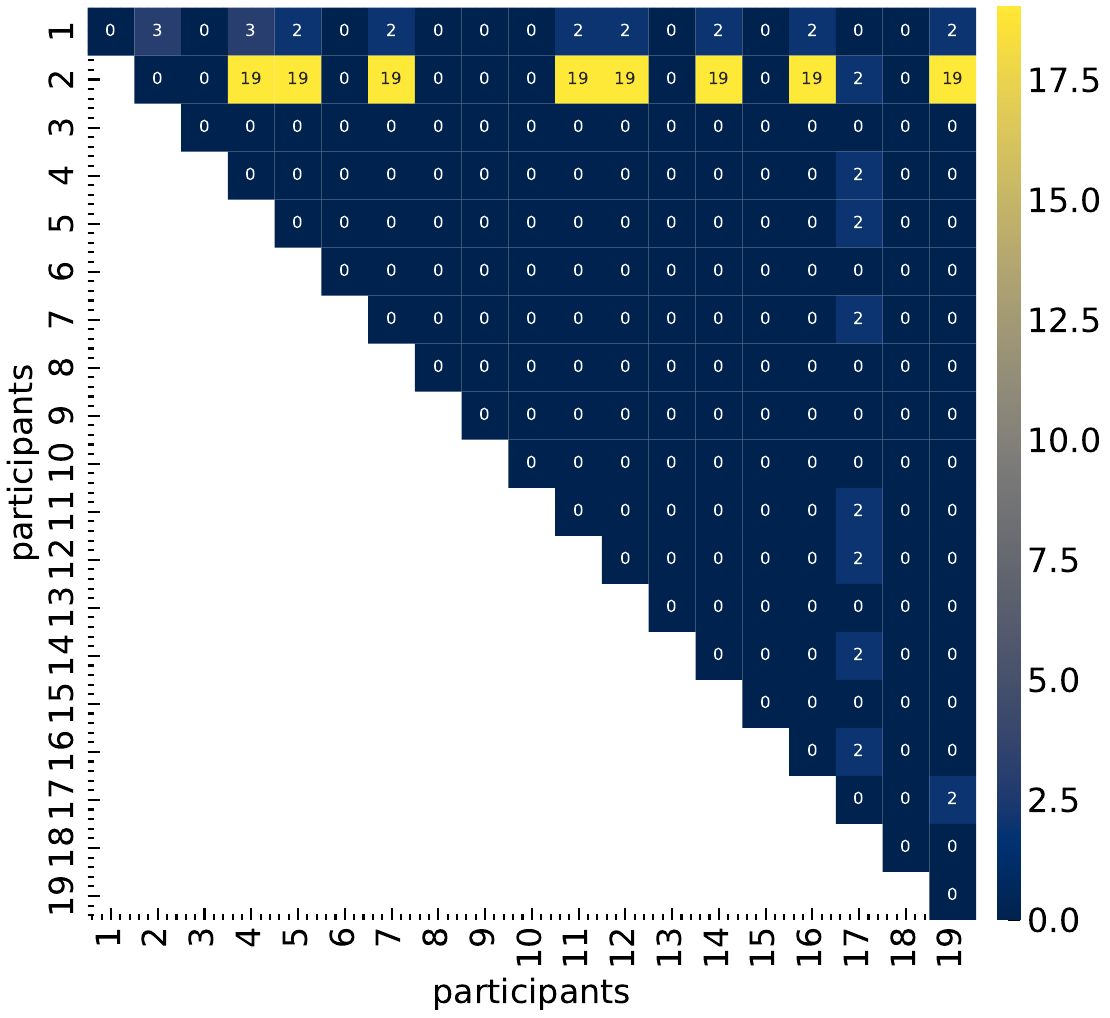}\label{fig:mozilla_remove_inprogress_overlap}}
     \subfloat[][Incogni]{\includegraphics[width=.25\linewidth]{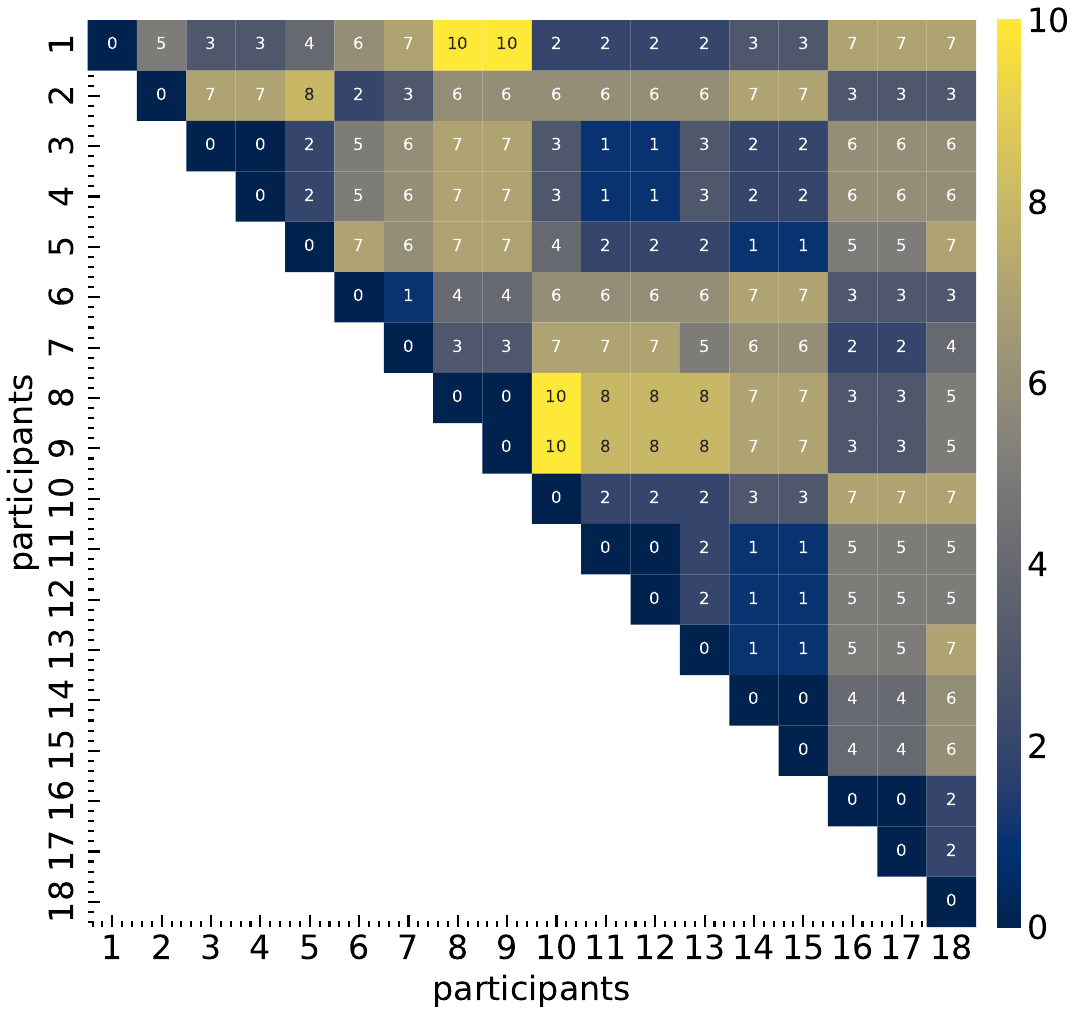}\label{fig:incogni_remove_inprogress_overlap}}
     \subfloat[][Kanary]{\includegraphics[width=.25\linewidth]{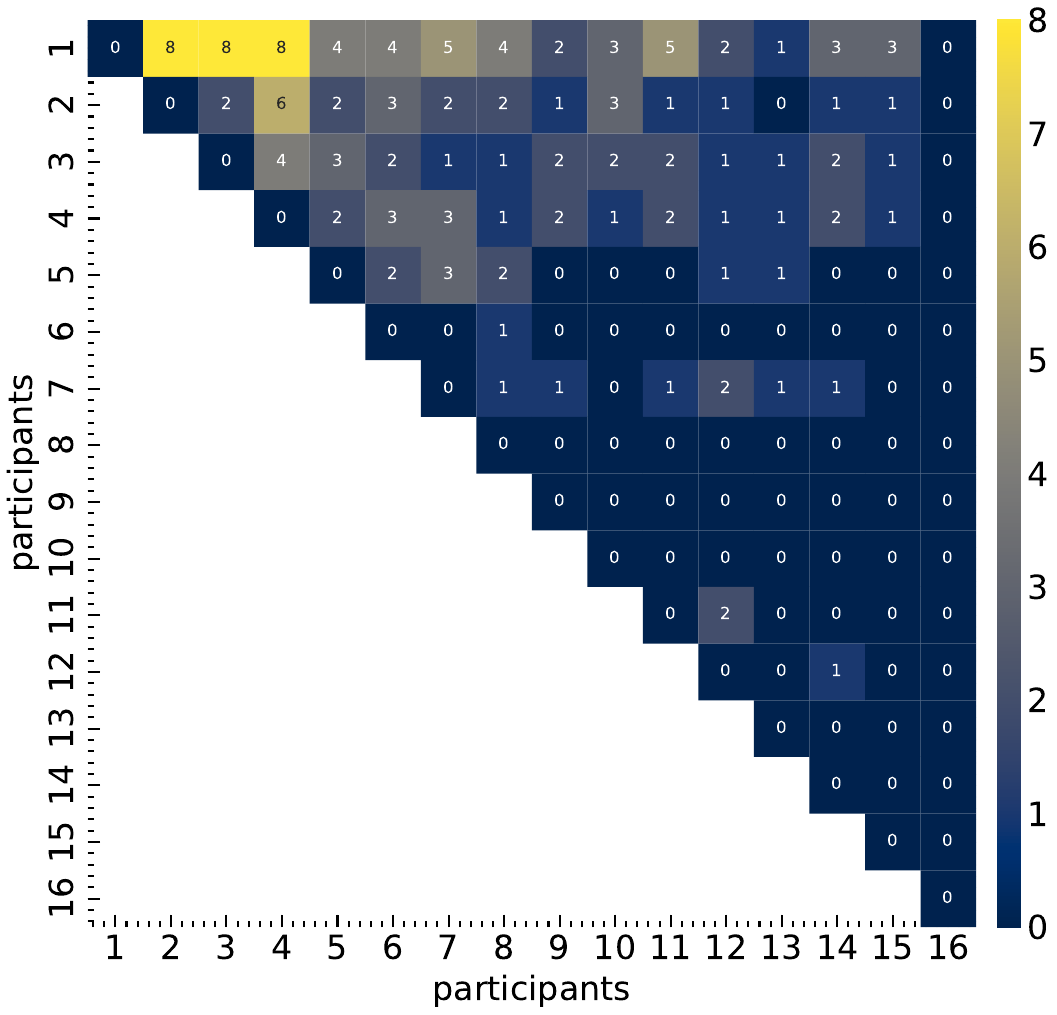}\label{fig:kanary_remove_inprogress_overlap}}
     \caption{Removal gap between each user of the PII removal service.}
     \label{fig:inside_efficiency}
\end{figure*}

The previous section examined the ability of removal services to detect records within their covered data brokers. However, simply finding a record does not guarantee its removal. Therefore, we finally evaluate the efficacy of the PII removal services in deleting the records they have previously identified. We investigate the extent to which these services can successfully remove the discovered PII, thereby providing insight into their practical effectiveness in enhancing user privacy.

\pb{Identifying Successful Removals.}
The removal services all provide a status code for each retrieved record to indicate whether it has been successfully removed from the data broker (collected via our plugin). Since the status codes of each service are slightly different, we manually consolidate all the status codes and categorize them into three main groups: 
\one \textbf{Removed / Not found}: The record has been successfully removed from the data broker, or participant's PII is not found in data broker; 
\two \textbf{In progress}: The service has submitted an opt-out request for the record to the data broker, but the data broker has not responded yet;
and
\three \textbf{Failed}: This record removal failed, possibly due to an internal server error that prevented a request from being sent to the data broker, or the data broker did not respond for a long time after receiving the request.

\pb{Overall Efficacy.}
Figure \ref{fig:removal_result} shows the removal status of each participant in each service after 30 days of subscription. Overall, the four removal services in our experiments are relatively inefficient. Over the period of a one-month subscription, only an average of 48.2\% of each participants' records are successfully removed. Of these, \texttt{Incogni} has the highest successful removal rate, with an average of 76.6\% of records removed per participant. The least effective is \texttt{Kanary}, with an average of only 23.4\% of records removed per participant.

\pb{Per user success for the same service.}
We observe that there is a difference in removal efficacy among different users of the same service. For a given removal service, one user's records might be successfully removed from a broker, whereas another user's are not. To quantify the gap, we calculate the number of intersections between what has been removed and what hasn't to reflect the value of removal gap per pair of users. Thus, we define the removal efficacy gap between $\text{User A}$ and $\text{User B}$ as: 
\begin{align*}
    \text{\{A's removed brokers\}} \cap \text{\{B's in progress \& failed brokers\}}\; +\; \\
    \text{\{B's removed brokers\}} \cap \text{\{A's in progress \& failed brokers\}}
\end{align*}

Figure \ref{fig:inside_efficiency} shows heatmaps of the removal efficacy gap between all users of each service. We confirm that, for different users of the same service, one user's records may have been removed, while another user's records from the same data broker is still in progress.
Indeed, all services have cases where there is a removal efficacy gap among users. 
The largest gap between users is with \texttt{Optery}, which has an average gap of 72.5 data brokers per pair of users, \ie for each user, on average, \texttt{Optery} successfully removed data from 72.5 brokers, while simultaneously failing to remove data from that same broker for its other users.
The removal efficacy gap between users in the other three services is relatively small though, with average values of 4.5 (\texttt{Incogni}), 1.2 (\texttt{Kanary}), and 1.1 (\texttt{Mozilla Monitor}).

This indicates that even with the same subscription to the same PII removal service, there may still be gaps in removal efficacy, as perceived by different users. One potential explanation is that the records retrieved by the service may contain varying amounts of PII for different users (\ie one records may contain only phone numbers and addresses, while another may contain more PII such as family relationships, dates of birth), leading to differences in the difficulty removing records.

\begin{figure}
     \centering
     \includegraphics[width=0.6\linewidth]{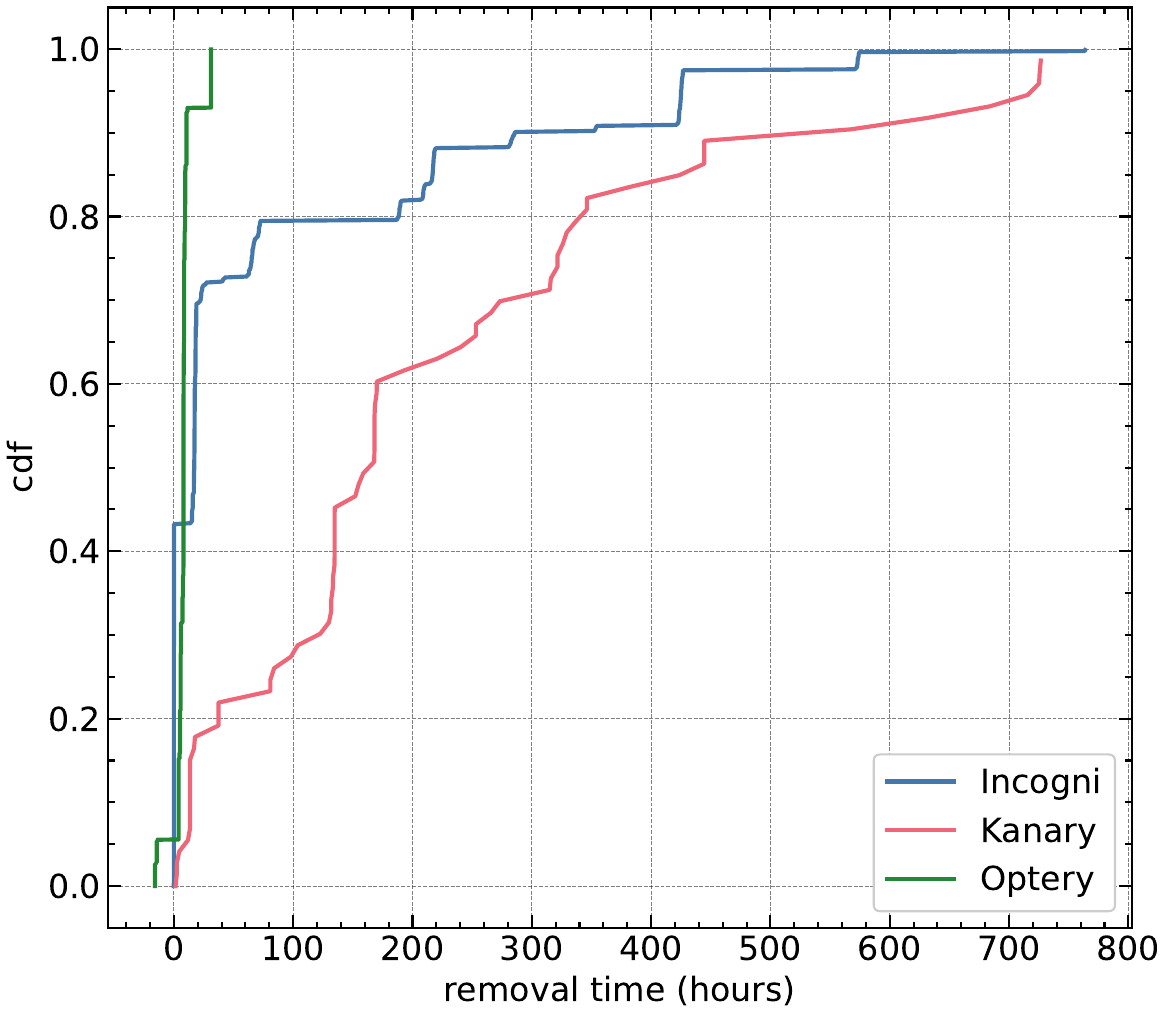}
     \caption{CDFs of removal times for all removed records for each PII removal service.}
     \label{fig:removal_time_cdf}
\end{figure}

\begin{figure*}
     \centering
     \subfloat[][]{\includegraphics[width=.38\linewidth]{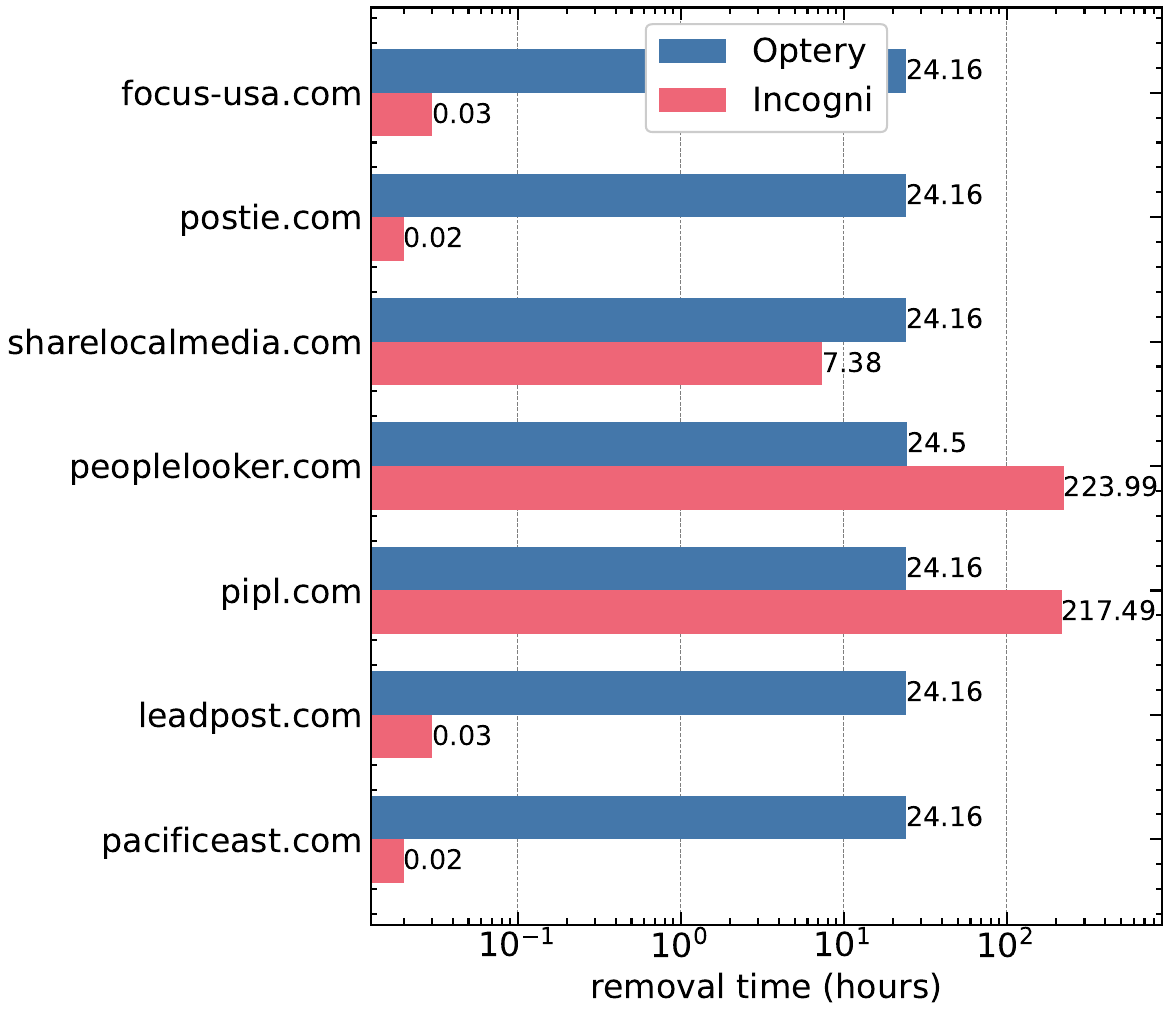}\label{fig:optery_vs_incogni}}
     \subfloat[][]{\includegraphics[width=.38\linewidth]{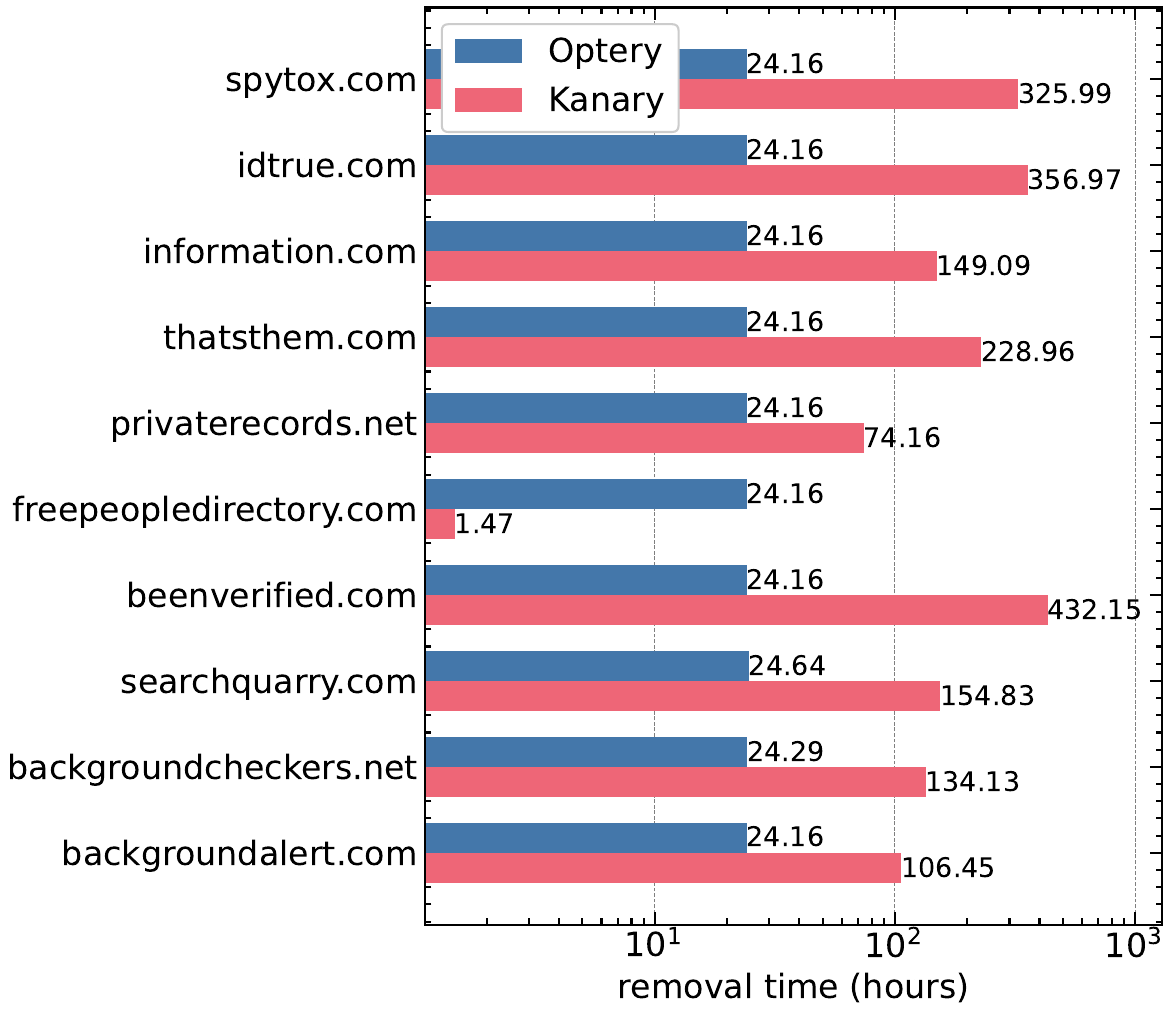}\label{fig:optery_vs_kanary}}
     \caption{(a) Comparison of \texttt{Optery} (blue) and \texttt{Incogni} (red) removal times for the same data brokers (note log scale on X-axis). (b) Comparison of \texttt{Optery} (blue) and \texttt{Kanary} (red) removal times for the same data brokers (note log scale on X-axis).}
\end{figure*}

\pb{Removal Delay.}
We next examine the time taken by different services to remove PII. We argue that faster removals reduce the risk of user PII exposure. All services except \texttt{Mozilla Monitor} provide the time the record was retrieved, and the time the record was removed.
Figure \ref{fig:removal_time_cdf} shows a CDF plot of the time elapsed from retrieval to removal for each service. 

Overall, \texttt{Optery}'s removal is significantly faster, with all removals occurring within the first 32 hours of the subscription. However, there are no other updates for the remainder of the one-month subscription.
Similarly, \texttt{Incogni} removes the majority of records at the beginning of the subscription period, with 72.2\% of removals occurring in the first 32 hours of the subscription. 
\texttt{Kanary}'s removal process is relatively slower, with only 19.2\% of records being removed within the first 32 hours of the subscription.
This suggests that there is a significant difference in removal times across services.
This may be due to the fact that different data brokers have different levels of removal difficulty.

\pb{Comparison of Removal Time on Shared Brokers.}
To explore this, we compare the removal times of different services when contacting the \emph{same} data broker. 
Note, there is no common data broker accessed by both \texttt{Incogni} and \texttt{Kanary}. 

Figure \ref{fig:optery_vs_incogni} and Figure \ref{fig:optery_vs_kanary} show the average removal time of services on the same data broker. 
Overall, there is a significant difference in removal times across services for the same data broker. \texttt{Optery}'s removal times are more stable, whereas \texttt{Kanary}'s removal times are generally much higher than those of Optery.
This is surprising as the mechanism to remove PII from a specific data broker is usually fixed (\eg by submitting a request on the website, or sending an email to the data broker).
Potential reasons for the differences in removal times include a lack of responsiveness from the data broker, or delays in the service's ability to update its removal status promptly. 

\takehomebox{\textbf{Take homes}: 
\one The four removal services are relatively ineffective, with an average removal success rate of 48.2\%. \texttt{Incogni} is the best performer at 76.6\%, while \texttt{Kanary} is the worst at 23.4\%. \two There is a gap in removal effectiveness among users of the same service, with \texttt{Optery} showing the largest gap of 72.5 between users. 
\three \texttt{Optery} has the shortest removal time, completing all removals in 32 hours.
\four There is a disparity in removal delays across different services even for the \emph{same} data broker, with \texttt{Optery} showing more stable delay and \texttt{Kanary} generally taking longer.}

%% file: Section/7.Discussion.tex
\section{Discussion and Implications}
\label{sec:discussion}

Our study has revealed a number of issues with the current data broker ecosystem. We next discuss other aspects worth studying, forming our future work.

\subsection{Discussion \& Limitation}

\pb{Impact of User Study Demographics on Generalizability.}
First, we note that the demographics of our user-study participants may differ from the demographics of the users of PII removal services. This may impact the generalizability of our results.

The participants in our user study are mainly students from a university in the United States. As a result, the demographic characteristics of the participants likely differ significantly from the demographic characteristics of the people who use the PII removal services we studied. As some examples, compared to the ``median'' user of a commercial PII removal service, we expect that the participants in our study to be (on average): \one~younger, \two~have shorter employment histories, \three~to be less likely to own property, and \four~less likely to have been involved in court filings, \etc
These (potential) demographic differences could impact the types and amounts of PII that data brokers hold about a person, and so indirectly the accuracy and amount of information that a PII removal service could remove.

We cannot know for certain how the demographics of our study compare to the demographics of each service's user base (\texttt{Mozilla Monitor}, \texttt{Optery}, \texttt{Kanary}, and \texttt{Incogni} do not publish demographic information about their users). Nevertheless, we flag the possibility (or likelihood) because of the potential impact on the generalizability of our results. More broadly, we emphasize that it is possible that PII removal services work better, worse, or just otherwise differently for the typical \Web{} users than they do for students. Readers should therefore consider our results accordingly.

\pb{Impact of Selected Services on Generalizability.}
Second, we note that our user study was limited to \NumServicesDetailedWord{} PII removal services, and that this limitation affects how generalizable our findings are. For a variety of reasons, we were unable to study all existing commercial PII removal services (\EG{} budget limitations, limited number of participants in our user study, complexity of adding support for each new service in our browser extension). We selected the \NumServicesDetailedWord{} services with the intent of capturing a representative sample of the industry. In some cases this is because the services are popular and claim to have large user bases (\ie \texttt{DeleteMe}, \texttt{Mozilla Monitor}); in other cases because the selected services seem to share similar underlying implementations to other operating services (\ie \texttt{Optery}, \texttt{Kanary}).

Nevertheless, its possible that, despite our best efforts, the PII removal services we selected do not generalize to all companies in the field. There could be services we did not measure that perform significantly better (or worse) than the services included in our study. We encourage the reader to interpret our results with this limitation in mind, and note that a broader study, covering more services, would be useful future work.

\pb{Impact of PII Provided to Removal Services.}
Third, we note that for ethical reason, we do \emph{not} ask participants to record what PII values they input to the removal service (during their subscription). However, the specific PII required varies between each removal service, ranging from 4 to 10 items (in fact, only 3 types of PII are mandatory for every service, see Table \ref{table:required_pii}). This therefore introduces a variable that may impact the efficacy of the removal services. Examining which PII is most useful for the removal services in discovering records from data brokers would be useful future work, and could provide guidance for improving the services.

\pb{Ground-truth of Removals.} 
Finally, we note that due to ethical and budgetary constraints, we do not ask participants to verify whether their PII had been removed from data brokers as claimed by the removal services. One of the authors is subscribed to the four removal services evaluated (\texttt{Optery}, \texttt{Incogni}, \texttt{Kanary} and \texttt{Mozilla Monitor}) in this study. Through this, we confirmed that the PII was indeed removed from data brokers, as claimed. That said, we do not rule out the possibility of other removal services making false claims, and assessing their ``honesty'' would be a valuable direction for future work.

\subsection{Implications}

Our study has a number of key implications for both users and the wider industry, which we discuss next. 

\pb{Implications for Users.}
For users, these results serve as a reminder of the limitations of PII removal services. Despite the claims on their official website, our experiments find that the average successful removal rate is only 48.2\% per user.
Furthermore, we find that the overlap between the data broker coverage of the different PII removal services is low, with average Jaccard similarity of 0.21 (see \S\ref{subsec:service_characterization1}). With this in mind, we conjecture that users may benefit from using multiple removal services to enhance their chances of effectively removing their personal information. Building such services that automate this would be valuable. 

\pb{Implications for Industry.}
For the PII removal services industry as a whole, these findings emphasize the urgent need for improved standards and more effective solutions. 
We argue that removal service providers should particularly prioritize enhancing the accuracy of data deletion procedures.
Better enforcement of regulation may be crucial here, as we found that the majority (71.7\%) of data brokers are not listed in any of the four US states' registrars.
As such, efforts should focus on establishing clearer guidelines for PII removal procedures, ideally standardizing opt-out APIs.

%% file: Section/8.Related_Work.tex
\section{Related Work}
\label{sec:related_work}

\subsection{Studies of Data Brokers}
\label{subsec:related_work1}

Many prior efforts have analyzed the hazards of data brokers and the lack of data broker transparency.
Crain \cite{crain2018limits} examines the inherent challenges in achieving transparency within the data broker industry. It concludes that the commoditization of personal data by brokers seriously undermines the right to privacy and that stronger regulatory interventions are necessary.
Similarly, Pinchot \emph{et al.} \cite{pinchot2018data} explore various privacy issues associated with the data broker industry, arguing that the widespread collection and sale of PII data by brokers exacerbates privacy risks, emphasizing transparency of data brokers as a key issue and highlighting the need for stronger legal and ethical safeguards.
Rostow \cite{rostow2017happens} investigates the unique privacy risks that arise when a familiar person purchases PII data through a data broker, arguing that such transactions can lead to unexpected and potentially harmful privacy violations.
Abad \emph{et al.} \cite{abad2016social} examines data brokers' practices of collecting, analyzing, and selling PII on social networks. They find that data brokers collect large amounts of user data through user interactions, preferences, and shared content on social networks. They conclude that data brokers conceal from users how PII data is collected and the purposes for which it is collected. They further argue that data brokers exploit legal loopholes to carry out other activities with PII, which emphasizes the need for stricter regulation of data brokers.

A common aspect of these studies is that they highlight the lack of transparency in the practices of data brokers and the resulting damage to PII. However, while these studies provide valuable insights into systemic issues within the data brokerage industry, they focus primarily on the regulatory and ethical aspects of data brokers and do not involve large-scale analysis of brokers. In our study, we present the first large-scale collection of existing data brokers and perform an empirical analysis.

\subsection{PII Removal from Data Brokers}
\label{subsec:related_work2}

A small set of prior efforts have studied the removal of PII from data brokers.
Grauer \cite{Grauer_2024} recruit 32 participants to explore the efficiency of seven removal services. The results are similar to the findings of our work that removal services were largely ineffective, removing only 35\% of personal data profiles on average, with manual opt-outs performing better at 70\% removal but still incomplete. However, the small number of participants per removal service in this report limits the generalizability of the results.
Take \emph{et al.} \cite{take2022feels} explore the challenges users face in attempting to remove their personal information from people search websites, highlighting the persistent and often frustrating nature of this process, where data reappears despite removal efforts.
In another work, user privacy rights across 20 people search websites is explored \cite{take2024expect}. The authors find that most sites do not comply with data access requests. The study also highlights that removing data from certain sites can lead to removal from connected sites, suggesting that understanding these connections can streamline data removal.
Similarly, Habib \emph{et al.} \cite{habib2019empirical} investigate the challenges users encounter with data deletion and opt-out processes across various websites. It highlights significant inconsistencies and barriers, illustrating the lack of standardization that makes managing personal data privacy difficult. These findings underscore the necessity for improved regulatory frameworks and more accessible data management options, aiming to enhance privacy protection and simplify user experiences in the digital landscape.

To date, these works offer only small-scale studies of information removal from people searching sites. Critically, they do not involve a large number of data brokers or PII removal services. To the best of our knowledge, we offer the first large-scale study of PII removal services and data brokers. 

%% file: Section/9.Conclusion.tex
\section{Conclusion}
\label{sec:conclusion}

This paper has presented the first empirical study of PII removal services. We initially surveyed 10 major PII removal services and the 2,024 data brokers they cover. We discovered the small overlap (average Jaccard similarity of 0.21) in data broker coverage between these services, as well as the lack of corresponding regulation.
To evaluate the efficacy of such removal services, we then focused on four services, recruiting 71 participants to use them.
We found that that these PII removal services struggle to discover records accurately, with only 41.1\% of the user records being correct, and only 48.2\% of records successfully removed from data brokers. 
As discussed in \S\ref{sec:discussion}, there are many avenues of future work in this understudied area. We are particularly keen to explore the impact that participant demographics have on the efficacy of a wider set of PII removal services. It would also be valuable to develop free alternatives that streamline data removals for individuals. We hope our work can provide valuable insights to researchers, catalyzing such work.